\documentclass[aps,pra,twocolumn,superscriptaddress]{revtex4}

\usepackage{graphicx}
\usepackage{amsmath}
\usepackage{amssymb}

\begin{document}

%

\title{Band structure, elementary excitations, and stability of a
Bose-Einstein condensate in a periodic
potential}

\author{M. Machholm}
\email[]{machholm@nordita.dk}
\altaffiliation[Also at]{ {\O}rsted Laboratory, H. C. {\O}rsted
Institute, Universitetsparken 5,
     DK-2100 Copenhagen \O, Denmark}
\author{C. J. Pethick}
\email[]{pethick@nordita.dk}
\affiliation{NORDITA, Blegdamsvej 17, DK-2100 Copenhagen \O, Denmark}

\author{H. Smith}
\affiliation{{\O}rsted Laboratory, H. C. {\O}rsted Institute,
Universitetsparken 5,
DK-2100 Copenhagen \O, Denmark}




\date{\today}

\begin{abstract}
We investigate the band structure of a Bose-Einstein condensate in a
one-dimensional periodic potential by calculating stationary
solutions of the Gross-Pitaevskii equation which have the form
of Bloch waves. We demonstrate that loops (``swallow tails") in the band
structure occur both at the
Brillouin zone boundary and at the center of the zone, and they are
therefore a generic feature.
A physical interpretation of the swallow tails in terms of periodic
solitons is given.
The
linear stability of the solutions is investigated as a function
of the strength of the
mean-field interaction, the magnitude of the periodic potential, and the
wave vector of the condensate. The regions of
energetic and
dynamical stability are
identified by considering the behavior of the Gross-Pitaevskii energy
functional for small deviations of the condensate wave function from a
stationary state. It is also shown how for long-wavelength
disturbances the
stability criteria may be obtained within a hydrodynamic approach.
\end{abstract}



\maketitle

\section{Introduction}

The possibility of studying experimentally the properties of
quantum atomic gases in the periodic potential created by
interference of two laser beams, a so-called optical lattice, has
led to an explosion of activity, both experimental and
theoretical.  One of the basic properties of an atom moving in a
periodic potential is that it undergoes Bloch oscillations when
subjected to a sufficiently weak external force. In a gas of thermal
cesium atoms such oscillations were observed as long ago as 1996
\cite{salomon}.
Subsequently, a number of experimental studies of
Bose-Einstein condensates in optical lattices have  been made. In
one-dimensional lattices, interference has been observed between
condensates initially trapped in different local minima of the
potential \cite{kasevich}. Also Bloch oscillations~\cite{firenze1} and
Josephson oscillations~\cite{firenze2} of a condensate
have been observed, and acceleration and
collective behavior of a condensate have been studied \cite{pisa}.
In higher-dimensional lattices, interference effects have been
investigated and the transition to an insulating state has been
observed \cite{munich}.

On the theoretical side, Bloch oscillations have been investigated
\cite{kirstine}, and states of uniform flow have been studied
\cite{choi,WN1}. For sufficiently weak interparticle interaction,
the properties of a Bose-Einstein condensate resemble those of a
single particle moving in a periodic potential. Properties of a
condensate in this regime have been explored in Ref.~\cite{WN2}.
One of the surprising discoveries is that the interaction between
particles can influence the band structure dramatically.  In a
two-state model, Wu and Niu found a loop in the band structure at
the boundary of the first Brillouin zone \cite{WN1} and, in
detailed calculations of band structure, Wu {\it et al.} found evidence
for non-analytic behavior at the zone boundary \cite{WDN}. Another
unexpected discovery was that for sufficiently strong particle
interactions, there exists a simple exact solution to the
Gross-Pitaevskii equation for a condensate with a wave vector
$k$ corresponding to the boundary of the first Brillouin zone
\cite{bronski1, bronski2}. More recently Diakonov {\it et al.} have
carried out explicit numerical calculations of the band structure,
and have demonstrated that the band has a swallow-tail
feature at the zone boundary \cite{Diakonov2002}.  They also
showed that this behavior is predicted by the simple
two-component model used previously  by Wu and Niu \cite{WN1}.

One purpose of this article is to calculate properties of
stationary states of a Bose-Einstein condensate in an
optical lattice. We carry out numerical calculations of the
band structure and investigate the size of the loop at
the boundary of the first Brillouin zone. In addition, it is
demonstrated that a similar swallow-tail structure can arise at the
zone center. It is remarkable that the
width of the swallow tails remains nonzero even in the absence of the
lattice potential.  We show how this may be understood in terms of
periodic solitons which, for the Gross-Pitaevskii equation, were first
investigated by Tsuzuki \cite{tsuzuki}.

A second purpose of the paper is to explore elementary excitations
and stability of states of uniform superflow.  For relatively weak
interparticle interactions, this has been done in Refs.~\cite{WN1}
and \cite{WN2}, and we
shall in this paper devote most attention to the range of parameters for
which
there are loops in the band structure. Quite generally, linear
stability of states may be investigated by expanding the
Gross-Pitaevskii
energy functional to second order in the deviation of the condensate
wave
function from the solution for a stationary state. One may distinguish
two types of
stability.  The first is energetic stability, which is referred to in
Ref.~\cite{WN2} as ``Landau stability", and the condition for this
is that
the changes in the Gross-Pitaevskii energy functional due to the change
in the condensate wave function be
positive definite. The other form of stability is dynamical
stability, and the criterion for this is that
the linearized time-dependent Gross-Pitaevskii equation have no
complex
eigenvalues.
To describe
excitations with wavelengths long compared with the period of the
lattice, a hydrodynamic approach may be applied.
This work
represents a generalization to moving condensates of the
calculations of Ref.~\cite{kraemer} for condensates at rest. It
yields a
stability
criterion  for creation of long-wavelength phonons which reduces for a
translationally-invariant system to
the Landau criterion.

The results obtained in this work are derived from numerical solutions
to the one-dimensional Gross-Pitaevskii equation. The method we adopt is
to expand
the wave function in a Fourier series.
In order to elucidate
the physical meaning of our
results, we have also carried out approximate analytic calculations
which
yield simple results in qualitative agreement with those of
the
numerical calculations.

In this paper we shall consider only extended states having the form of
Bloch waves. Due to the nonlinear nature of the Gross-Pitaevskii
equation there are also stationary states corresponding to
localized
excitations such as isolated solitons~\cite{TS}, as well as states
in which the density varies periodically with a period different from that
of the optical potential.

This paper is organized as follows. In Sec.\ \ref{SecBW} the properties
of stationary states are described.  There we present analytical
and numerical calculations, and describe how swallow-tail structures
may be understood in terms of periodic solitons.  Section \ref{SecElex}
gives a general discussion of elementary excitations of a
condensate, and energetic
and dynamical stability.  It
describes the hydrodynamic approach applicable at long wavelengths.
Numerical results for the stability of a condensate in a one-dimensional
optical lattice are given in Sec.\ \ref{Stab1D}.
Section \ref{dis} contains a discussion of our results and concluding
remarks.

\section{Bloch waves}\label{SecBW}

The basic assumption that we shall make in this paper is that
fluctuation effects are so small that the state of the condensate
may be calculated in the Gross-Pitaevskii approach, in terms of
the condensate wave function $\psi({\bf r})$.  The energy of the
state is then given by
\begin{eqnarray}
E[\psi] = \int d{\bf r}
\left(\frac{\hbar^2}{2m}|\boldsymbol{\nabla}\psi|^2+V({\bf r})|\psi|^2
      +\frac{1}{2}U_0|\psi|^4\right).
\nonumber\\
      \label{energy}
\end{eqnarray}
Here $m$ is the mass of a particle, $V({\bf r})$ is the external
potential and $U_0$ is the effective interaction between two
atoms, which is given in terms of the scattering length $a$ for
two-body collisions by $U_0=4\pi \hbar^2a/m$.

Stationary states of the condensate may be found in the usual way
by demanding that the energy functional (\ref{energy}) be
stationary under variations of $\psi({\bf r})$, subject to the
condition that the total number of particles remain unchanged. This
yields the time-independent Gross-Pitaevskii equation
\begin{equation} -\frac{\hbar^2}{2m}\nabla^2\psi+V({\bf r})\psi
+U_0|\psi|^2\psi =\mu\psi, \label{gp}
\end{equation}
where $\mu$
is the chemical potential.

A one-dimensional optical lattice gives rise to a potential acting
on an atom which has the form
\begin{equation} V(x)=2V_0\cos^2(\pi
x/d)=V_0\cos(2\pi x/d) +V_0,
\label{potential}
\end{equation}
    where $d$ is the period
of the lattice.  The coefficient $V_0$, which measures the
strength of the potential, depends on the polarizability of the
atom and the intensity of the radiation that
generates the optical lattice. In future we shall generally neglect the
constant term, and take the potential to be simply $V_0\cos(2\pi
x/d)$ \footnote{This choice agrees with that made in
Refs.~\cite{WN2} and \cite{WDN}, but is different from the
one made in Ref.~\cite{Diakonov2002}}. In addition, we shall
not take the potential due to the trap into account. Such an approach
should give a good approximation to the local properties of stationary
states of the condensate,
provided
the average density and average wave number of the condensate vary
slowly in
space on the scale of the period of the optical lattice.  For studying
excitations, this approach will be valid provided the
wavelength of the
excitations is large compared with the lattice spacing but small
compared with the distance over which properties of the unperturbed
condensate vary significantly.
A further assumption we shall make is
that the states are uniform in the $y$ and $z$ directions. The resulting
Gross-Pitaevskii equation has a variety of
different sorts of stationary solution.  Some of these are extended,
while
others, such as solitons, are localized in space. In this paper we
shall focus on extended solutions to
Eq.\ (\ref{gp}). These are the analogs of Bloch states for a single
particle in a lattice.  As remarked in the introduction,
there are stationary solutions of
the
Gross-Pitaevskii equation for which the particle density does not
have the same period as the lattice.
As will be explained elsewhere, they
are related to the self-trapped states of a condensate in a
double-well potential \cite{milburn, smerzi, coullet,mahmud}. Here we
shall confine our attention to solutions of the usual Bloch form
\begin{equation}
\label{Blochwave}
\psi(x)={\rm e}^{ikx}f(x) ,
\end{equation}
where $\hbar k$ is the quasimomentum and $f(x)$ has the same period as
the lattice, $f(x)=f(x+d)$.

The energy per
unit volume, ${\tilde E}$, is then given by
\begin{eqnarray}
{\tilde E}&=&\nonumber
\frac{1}{d}\int_{-d/2}^{d/2}dx
\left[\frac{\hbar^2}{2m}\left|\frac{d
\psi}{dx}\right|^2\right.\\
&&\left. +V_0\cos
\left(\frac{2\pi x}{d}\right)|\psi|^2
+\frac{1}{2}U_0|\psi|^4\right].\label{energy1}
\end{eqnarray}

We determine equilibrium solutions $\psi$ of the time-independent
Gross-Pitaevskii equation by expanding
$\psi$ in plane waves,
     \begin{equation} \psi_k= \sqrt{n}{\rm e}^{ikx}\sum^{\nu_{\rm
max}}_{\nu=-\nu_{\rm
max}}
     {a_{\nu}{\rm e}^{i2\pi\nu x/d}},\label{expansion} \end{equation}
where $\nu$ is an integer. Here
\begin{equation}
       n=  \frac{1}{d}\int_{-d/2}^{d/2}dx |\psi|^2
\end{equation}
is the average particle density.  From this it follows that
the coefficients $a_{\nu}$ satisfy the normalization condition
\begin{equation}
\sum^{\nu_{\rm max}}_{\nu =-\nu_{\rm max}}|a_{\nu}|^2    =1
.\label{norm2}
\end{equation}

The stationary states of the system may be obtained by a variational
method, by requiring that the derivatives of the energy
functional (\ref{energy1}) with respect to
$a_{\nu}$ vanish.  There are $2\nu_{\rm max}+1$ complex variables, and
one real
constraint (\ref{norm2}).  In addition, the overall phase of the
wave function is arbitrary, so the energy functional depends on
$4\nu_{\rm max}$ independent real variables.

A considerable simplification of the computational effort is possible
because it turns out that for the stationary states of interest for the
range of parameters we have considered, the phases $\phi_{\nu}$ of the
coefficients $a_{\nu}$ may be taken to be the same, or to differ by
$\pi$.    It is easy to demonstrate that such
states are indeed stationary  under variations of the phases because the
phases occur in the energy  functional as terms of the
type $\cos(\phi_{\nu_1} -\phi_{\nu_2})$ and $\cos(\phi_{\nu_1}
+\phi_{\nu_2} -\phi_{\nu_3} -\phi_{\nu_4})$.  The derivatives of these
functions with respect to the phases thus give terms of the type
$\sin(\phi_{\nu_1} -\phi_{\nu_2})$ and $\sin(\phi_{\nu_1} +\phi_{\nu_2}
-\phi_{\nu_3} -\phi_{\nu_4})$, which vanish if the phases are zero or
$\pi$. We sought solutions with other phases by allowing the
coefficients  to be complex, but found none. Because the overall phase
is arbitrary, this means that we may
take the $a_{\nu}$ to be real. When this is done,  there are
only  $2\nu_{\rm max}$ independent real
variables.

We shall present our results in terms of the energy per particle, ${\tilde
E}/n$, as a function of the (one-dimensional) wave vector $k$. As a
convenient unit of energy we employ the quantity $E_0$ given by
\begin{equation}
E_0=\frac{\hbar^2\pi^2}{2 m d^2},
\end{equation}
which is the kinetic energy of a particle with wave vector
equal to that at the
boundary of the first Brillouin
zone. For an optical lattice made by oppositely directed laser
beams, the lattice spacing is half the wavelength $\lambda$ of the
light, and therefore $E_0$ is equal
to
the
kinetic
energy
given to an atom initially at rest when it absorbs a
photon
having the frequency of the laser.

An unusual feature of the resulting energy bands is the
appearance of loops in the form of swallow-tail structures, as
demonstrated in Ref.~\cite{Diakonov2002}. One noteworthy
result of the present work is that swallow tails can occur also at
the zone
center
in higher-lying bands, as
illustrated in Fig.~\ref{Bandstructure2}, which shows results of
numerical
calculations of the band structure that will be described in Sec.\
\ref{numerics}. Thus the
appearance of swallow tails is a general feature.
At the zone boundary, the swallow tail
appears when the interaction energy per particle $nU_0$ exceeds the
amplitude $V_0$ of the potential due to the optical lattice.
As the parameter $nU_0$ grows with respect to $V_0$, the
swallow tail increases in width and may extend deep into the zone.
The
condition $nU_0>V_0$ is necessary for the swallow-tail structure to
appear  at the zone boundary, $k=\pi/d$.

Had swallow
tails appeared only at the zone boundary, one might have suspected that
their existence was related to the fact that for $k=\pi/d$ there is
an exact solution to the Gross-Pitaevskii equation. However, for $k=0$
there is no exact solution of the Gross-Pitaevskii equation, and
consequently the existence of swallow tails is not connected with the
existence of an exact solution.
As we shall see
in detail below, the condition for the appearance of a loop at  the
zone center can be much less restrictive than at the zone boundary.
Before discussing our numerical results further, we shall now
analyze a simple model for the band structure near the zone center which
exhibits the main qualitative features of the full numerical
calculations.

\subsection{An analytic model}
\label{STstructure}

In Ref.~\cite{Diakonov2002} the swallow-tail structure of the
lowest energy band at the boundary of the first Brillouin zone was
discussed in terms of a simple trial solution to the
Gross-Pitaevskii equation. In order to illustrate the generic
nature of the phenomenon we shall here consider the band structure
at the center of the Brillouin zone, corresponding to $k=0$.

We
employ a trial function of the form
\begin{equation}
\psi=
\sqrt{n}{\rm e}^{ikx}(a_0 + a_1{\rm e}^{i2\pi x/d}+a_{-1}{\rm e}^{-i2\pi
x/d}),
\label{trial}
\end{equation}
where the coefficients $a_0$, $a_1$  and $a_{-1}$ are
chosen to be real for the reasons given above.
The trial function thus mixes into the free-particle wave function
$\exp(ikx)$ states that differ by the smallest reciprocal lattice
vectors, $\pm 2\pi/d$.  The normalization
condition (\ref{norm2}) is
\begin{equation}
a_0^2+a_1^2+a_{-1}^2=1.
\label{norm3}
\end{equation}
This constraint is satisfied automatically by expressing
the coefficients in terms of two angles $\theta$ and $\phi$ according
to the equations
\begin{equation}
a_0=\cos\theta, \;\;a_1=\sin\theta\cos\phi, \;\; {\rm and} \;\;
\;\;a_{-1}=\sin\theta\sin\phi. \label{trial1}
\end{equation}
Upon inserting the trial function (\ref{trial}) with coefficients
given by Eq.~(\ref{trial1}) into Eq.~(\ref{energy1}), we obtain
\begin{equation}
\frac{{\tilde E}}{n}=\epsilon_{\rm kin}+\epsilon_{\rm
pot}+\epsilon_{\rm int},\label{en1}
\end{equation}
where
\begin{eqnarray}
\epsilon_{\rm kin}= \frac{\hbar^2}{2m}\left[ k^2
+2k\left(\frac{2\pi}{d}\right)\sin^2\theta(\cos^2\phi-\sin^2\phi)\right.
\nonumber\\
\left. +\left(\frac{2\pi}{d}\right)^2
\sin^2\theta \right]
\label{en2}
\end{eqnarray}
is the kinetic energy,
\begin{equation}
\epsilon_{\rm pot}=V_0\sin\theta\cos\theta(\cos\phi+\sin\phi)
\label{en3}
\end{equation}
is the potential energy, and
\begin{eqnarray}
\epsilon_{\rm
int}=nU_0\left(\frac{1}{2}+\cos^2\theta\sin^2\theta(\cos\phi+\sin\phi)^2
\right.\nonumber\\
+\left.\frac{1}{4}\sin^4\theta\sin^2 2\phi\right) \label{en4}
\end{eqnarray}
is the interaction energy. The stationary points of
this energy function are obtained by equating to zero
the derivatives of the energy per particle
with
respect to $\theta$ and $\phi$. The results shown in
Fig.~\ref{Bandstructure} were found with this model.

\begin{figure}[t]
\includegraphics[width=80mm]{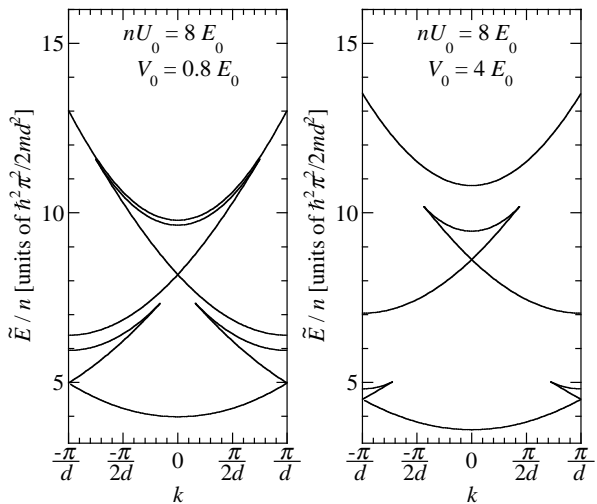}
\caption{\label{Bandstructure2} Energy per particle as a function of
wave number for the lowest bands. The results are obtained from
numerical calculations
based on the wave function (\ref{expansion}), as described in Sec.\
\ref{numerics}.}
\end{figure}

\begin{figure}[t]
\includegraphics[width=80mm]{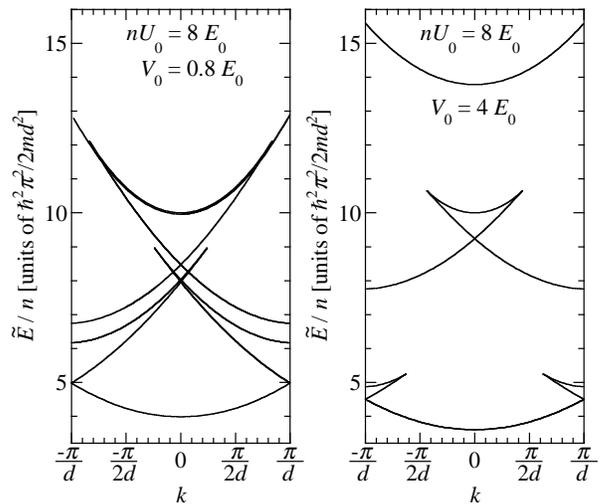}
\caption{\label{Bandstructure} Model calculations of the energy
per particle as a function of wave number for the same parameters as
were used
in Fig~\ref{Bandstructure2}.
The
results are
obtained using the variational method with the trial function
(\ref{trial}).}
\end{figure}

In order to exhibit the nature of
the solutions, we first consider the simple case in the absence of
interactions ($U_0=0$), and with $V_0\ll E_0$, so the lattice potential
is a small perturbation. At $k=0$ one finds three
stationary points with different energies. The lowest state corresponds
to the bottom of the lowest band, and its wave
function is a plane wave with $k=0$ plus a small admixture of states
with $k=\pm 2\pi/d$.  Its energy is given to second order in $V_0$ by
$E=-(V_0)^2/8E_0$.  The two other states are comprised primarily of
plane
waves with $k=\pm 2 \pi/d$ with a small admixture of the $k=0$ state.
The energies of the two states are $E=4E_0$ for the state which
corresponds to the top of the second band, and $E=4E_0+(V_0)^2/8E_0$ for
the state at the bottom of the third band.
Because of the simplicity of the trial function, there are no higher bands
in this model.
The magnitude of the energy gap at $k=0$ between the
second and third bands is thus $V_0^2/8E_0$. The reason that it is of
second order in the lattice potential is that the potential is
sinusoidal with period $d$.  Consequently, it couples
directly only states with wave numbers
differing by the smallest reciprocal lattice vectors, $\pm 2\pi/d$. The
coupling between states with wave vectors $k+2\pi/d$  and $k-2\pi/d$
is indirect, since it is brought about by the coupling of these states
to
the state with wave vector $k$. In the absence of interactions between
particles, the stationary points of the energy functional for $k=0$ have
$|a_1|=|a_{-1}|$, that is $\phi$ is an odd multiple of $\pi/4$.

\begin{figure}[t]
\includegraphics[width=80mm]{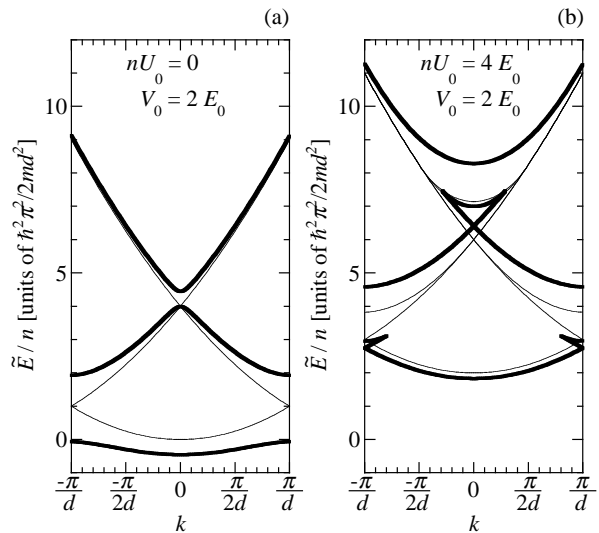}
\caption{\label{Bandstructure3} Energy per particle in the first
Brillouin zone as in Fig~\ref{Bandstructure}. The results are
obtained by a variational method with the trial function given in
Eq.~(\ref{trial}).
a) In the absence of interaction the band structure
(bold curves) exhibits the usual band gaps at $k=0$ and $k=\pi/d$. The
band gap is $V_0$ at $k=\pi/d$ and $V_0^2/8E_0$ at $k=0$
for small $V_0$. The thin curves show the energies for
$V_0 \rightarrow 0$, i.e. for the free non-interacting system. b) In the
presence of interaction the swallow tails appear for $U_0$ larger than a
critical value, which depends on $V_0$ and is different for the two band
gaps (bold curves). The thin curves illustrate the limit
$V_0
\rightarrow 0$.}
\end{figure}

Figure~\ref{Bandstructure3}a shows results for non-interacting particles
($U_0=0$). The familiar band structure is seen. The
band gap is $V_0$ at $k=\pi/d$ and $V_0^2/8E_0$ at $k=0$
for small $V_0$, which is still a good approximation at $V_0= 2E_0$. The
lowest band is pushed down in the presence of the periodic potential
\footnote{Had we used the original lattice potential given by Eq.\
(\ref{potential}), which contains a constant term, the energy per particle
would increase by $V_0$ and, consequently, the energy per particle would
be positive for small $V_0$.}.

In the presence of interactions ($U_0\neq 0$), the energy landscape
in the $\theta$-$\phi$ plane has a more complicated structure, and the
number of points at which the energy functional is stationary
can be greater than for $U_0=0$. Figure~\ref{Bandstructure3}b exhibits
the resulting band structure. The band gap
at
$k=0$ is enhanced by the interaction, in contrast to the band gap at
$k=\pi/d$, which is reduced.

Let us now examine the behavior of the band structure in the limit of
vanishing lattice potential ($V_0 \rightarrow 0$).
As one would expect, the band gaps tend to zero while, in contrast,
the widths of swallow tails increase
with decreasing $V_0$, and they are nonzero for $V_0\rightarrow 0$. In
this
limit, states corresponding to the upper edge of a swallow tail become
degenerate with states in the band above.  As we shall discuss in  Sec.\
\ref{soliton}, the reason for this surprising behavior is that states
on the upper edge of a swallow tail correspond to periodic
soliton solutions of the Gross-Pitaevskii equation.

The appearance of a swallow tail requires that the interaction energy
$nU_0$ be sufficiently large. In the calculations for
states near the zone boundary in Ref.\ \cite{Diakonov2002}, it was shown
that the condition for existence of the swallow tail is $nU_0 >V_0$. We
now carry out a similar calculation for the swallow tail at the zone
center, using the three-state model described above.
We investigate the form of the energy function
(\ref{en1}--\ref{en4}) for $k=0$ near the point
$\theta=\pi/2$, $\phi=3\pi/4$.  This is a stationary point both for
$U_0=0$ and for $U_0\neq 0$. If one imagines
that $U_0$ increases gradually from zero while the other parameters
remain fixed, one observes in the energy landscape given by
(\ref{en1}--\ref{en4}) that two additional stationary points are split off
from $(\theta, \phi)=(\pi/2,3\pi/4)$ when $nU_0$ exceeds a critical
value. This value may be found by inserting $\theta=\pi/2+\delta$ and
$\phi=3\pi/4+\epsilon$ into Eqs.~(\ref{en1}--\ref{en4}) and expanding
the
energy to second order in $\delta$ and $\epsilon$. The resulting
expression for the change in the energy per particle relative to its
value for $\delta=\epsilon=0$
is \begin{equation}
\frac{{\tilde E}}{n}- 4E_0-\frac{3}{4}nU_0\simeq
-4E_0\delta^2+\sqrt{2}V_0\delta\epsilon
-\frac{nU_0}{2}(\delta^2+2\epsilon^2). \label{quaform}
\end{equation}
For $nU_0 = 0$, the point
$\delta=\epsilon=0$ is a saddle point.
As $nU_0$
increases, this point turns into a local maximum, and two saddle points
move out to points with both $\delta$ and $\epsilon$ nonzero.
The condition for the saddle point to turn into a local maximum is
that the symmetric matrix yielding the quadratic form
(\ref{quaform})
have a zero eigenvalue, and this occurs for $nU_0(nU_0+8E_0)=V_0^2$ or
\begin{equation}
nU_0=[(4E_0)^2+V_0^2]^{1/2}-4E_0. \label{critc}
\end{equation}
For
values of the lattice potential small compared with $E_0$,
($V_0\ll E_0$), the
critical value obtained from Eq.\ (\ref{critc}) becomes
$nU_0=V_0^2/8E_0$.
This is
physically reasonable, since the magnitude of the energy gap in the
absence
of interactions equals $V_0^2/8E_0$, as discussed above. At the zone
boundary, however, the energy gap in the absence of interactions
is equal to $V_0$, corresponding to the condition $nU_0>V_0$ for the
appearance of the swallow-tail structure at $k=\pi/d$. Note that the
result (\ref{critc}) is derived from an approximate trial
function which becomes exact only in the limit $V_0\ll E_0$.  Even
so, it yields a reasonable description of the dependence of
the critical value of $nU_0$ on $V_0$ also for higher values of
$V_0$, as shown by the numerical
results for the width of the swallow tail in Fig.~\ref{Width0}
below.

The swallow-tail structure illustrated in
Figs.~\ref{Bandstructure2}-\ref{Bandstructure3} is thus a
general phenomenon, in that it occurs both at the zone boundary and at
the zone center, provided the interaction energy $nU_0$ is
sufficiently large.  The phenomenon also occurs at other points in
the Brillouin zone for solutions which have a
periodicity different from that of the optical lattice, as will be
demonstrated in future work.

\subsection{Numerical calculations of band structure}
\label{numerics}
At the beginning of this section we described the general method
used to calculate stationary states of a moving condensate. Our numerical
procedure is as follows: The trial function (\ref{expansion}) is
inserted into the energy functional (\ref{energy1}), and stationary
points of the resulting expression for the energy are found. The
normalization of the wave function is
imposed as in Eq.\ (\ref{norm2}), or by generalizing Eq.\ (\ref{trial1})
to hyperspherical coordinates.
We determine the solutions by using the
Mathematica$^{\circledR}$ routine ``FindRoot", which requires as
input
an initial guess for the solution. The latter is found by
first solving the problem in a reduced
basis. Once a solution has been found for a
particular value of $k$, this is used as the initial guess for
nearby values of $k$.

In Fig.~\ref{Bandstructure2} two examples of numerical results for the
band structure are presented. The results are calculated with
$\nu_{\rm max}=4$.
This corresponds to a number of basis functions
equal to $2\nu_{\rm max}+1$.
The results for $\nu_{\rm max}=4$ and $\nu_{\rm max}=5$ differ by less
than
the
thickness of the lines.
Around $k=\pi/d$ the energy of the second band calculated with
$\nu_{\rm max}=2$ differs by less than 1\% from the full numerical
result.
Around $k=0$ and for higher bands, more plane waves contribute, and
typically
$\nu_{\rm max}=3$ is needed for 1\% convergence.
Compared to the simple model used in
Fig.~\ref{Bandstructure}, the numerical results exhibit smaller
swallow tails. The higher-lying energy bands are shifted downwards more
than
the lower-lying ones, causing the band gaps to become narrower.

\begin{figure}[t]
\includegraphics[width=80mm]{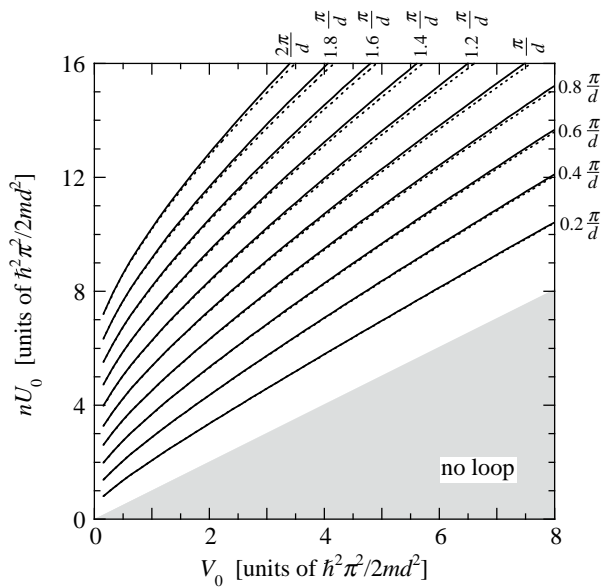}
\caption{\label{Width} Contour plot of the width of the swallow
tail in the lowest band around $k=\pi/d$.  The shaded area indicates the
region
where the swallow tail is absent, i.e., for $nU_0< V_0$.}
\end{figure}

\begin{figure}[t]
\includegraphics[width=80mm]{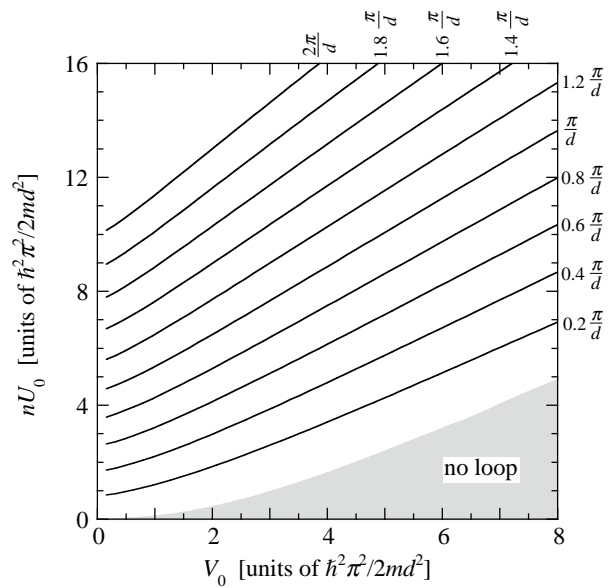}
\caption{\label{Width0} Contour plot of the width of the swallow tail in
the second band around $k=0$.
If the condensate wave function is approximated by the truncated
expression
(\ref{trial}), this swallow tail is absent if
the mean field
interaction is less than a critical value given by Eq.\
(\ref{critc}).  The corresponding region in parameter space is
indicated by the shaded area.}
\end{figure}

Rather than exhibiting the full band structure for different choices of
the  parameters, we choose to illustrate how one particular feature, the
widths of the swallow tails at the zone boundary and at the zone center,
depends on the two dimensionless quantities in the problem,
$nU_0/E_0$ and $V_0/E_0$. If the band structure is displayed in the
reduced zone scheme,
a swallow tail may be split up into a number of segments. This is shown
in Figs.\ \ref{Bandstructure2}-\ref{Bandstructure3}, where the swallow
tail
at
the zone boundary is divided into two halves. The full swallow tail may
be seen in an
extended zone
representation, and we define the width $w$ of a swallow tail as
being the magnitude of the difference between the
values of $k$ at the two tips of a swallow tail in this representation.
Widths of swallow tails are exhibited as contour
plots in Fig.~\ref{Width} and Fig.~\ref{Width0}.

In Fig.~\ref{Width} we show a contour plot of the width of the
swallow tail at the zone boundary in the first band as a function
of the mean-field interaction $nU_0$ and the potential parameter
$V_0$. The full lines are obtained from numerical solutions
with $\nu_{\rm max}=3$,  which is sufficiently large to ensure
an accuracy better than the thickness of the lines. The dotted lines
are results for $\nu_{\rm max}=2$.

The analytic model using the trial function (\ref{trial}) overestimates
the
width by less than 50\% for $nU_0 <8E_0$ and
$V_0<8E_0$. Table~\ref{tbl1} shows examples of calculated widths for
different basis sizes. We denote the widths calculated for a particular
value of $\nu_{\rm max}$ by $w_{\nu_{\rm max}}$. In all cases,
inclusion of more plane waves improves the result. The precision is
dependent  mainly on the width. Around $k=\pi/d$, fewer plane waves are
needed
than
around $k=0$ to describe the wave function to a given accuracy, and
therefore
the precision of the width estimates decreases with increasing width for
fixed $\nu_{\rm max}$. Correspondingly, the precision improves with
decreasing $nU_0$ and increasing $V_0$.

\begin{table}
\caption[t1]{Dependence on basis size of the calculated width $w_{\nu_{\rm
max}}$ of the swallow tail at the zone boundary. The table shows $w_6$ in
units of $\pi/d$, and the difference $w_{\nu_{\rm max}}/w_6 -1$ in per
cent.}
\label{tbl1}
\begin{tabular}{ll@{\ \ }|@{\ \ }c@{\ \ }|@{\ \ }c@{\ \ \ \ }c@{\ \ \ \
}c}
&&$$ & \multicolumn{3}{c}{Difference} \\
$V_0$ & $nU_0$ & $w_6$&\multicolumn{3}{c}{for $\nu_{\rm max}$}\\
&&& 1 & 2 & 3 \\
\hline
$E_0$ & $2E_0$ & 0.185547 &  13.6 & 0.25  &  0.00 \\
$E_0$ & $4E_0$ & 0.675781 &  27.1 & 1.05  &  0.02 \\
$E_0$ & $8E_0$ & 1.56555 &  46.7 & 3.19  &  0.13 \\
$4E_0$ & $8E_0$ & 0.552031 &  35.0 & 3.50  &  0.20 \\
$6E_0$ & $8E_0$ & 0.182891 &  27.8 & 2.69  &  0.13 \\
\end{tabular}
\end{table}

The width of the swallow tail in the second band around $k=0$ is shown in
Fig.~\ref{Width0}. As we described in the previous subsection, this
swallow tail
exists provided $nU_0$ exceeds a critical value, which for small values
of $V_0/E_0$ is given by $V_0^2/8E_0$. The analytic model using the
trial function
(\ref{trial}) estimates the width within 25\% for $nU_0/E_0<8$ and
$V_0/E_0<8$. Table~\ref{tbl2} shows some examples of the precision
within that parameter range.
In this case, the calculated widths using $\nu_{\rm max}=1$ are closer to
the full numerical results than are those calculated using $\nu_{\rm
max}=2$.
For $\nu_{\rm max}>2$ the precision systematically improves.
The precision becomes worse as 
$nU_0$ and $V_0$ increase.

The asymptotic behavior of the contours in Fig.~\ref{Width0} for $V_0
\rightarrow 0$ can be determined analytically within the simple model
described in
Sec.~\ref{STstructure}. The thin lines in Fig.\ \ref{Bandstructure3}
show an example of the band structure in this limit.
For fixed $nU_0$, the width of the
swallow tail increases
with decreasing $V_0$, and it is nonzero for $V_0=0$. At
the tip of the swallow tail these stationary points of the energy function
(\ref{en1}) merge at
$(\theta,\phi)=(\pi/2,0)$ for $V_0 \rightarrow 0$. The
interaction energy $\epsilon_{\rm int}$ has a minimum while the kinetic
energy
$\epsilon_{\rm kin}$ has a maximum at
$(\pi/2,0)$. Therefore the stationary
points merge when the upward curvature of $\epsilon_{\rm int}$
and the downward curvature of $\epsilon_{\rm kin}$ in the
$\phi$-direction cancel at $(\pi/2,0)$:
\begin{equation}
\left.\frac{\partial^2}{\partial\phi^2}\epsilon_{\rm
kin}\right|_{\theta=\pi/2}=-\left.\frac{\partial^2}{\partial\phi^2}\epsilon_{\rm
int}\right|_{\theta=\pi/2}.
\end{equation}
 From this criterion we obtain the asymptotic width of the
swallow tail at the zone center $w=(\pi/4d)nU_0/E_0$ for $V_0
\rightarrow 0$. Compared
to the numerical result in Fig.~\ref{Width0}, this analytic result
underestimates the asymptotic value of $nU_0$ for $V_0 \rightarrow 0$ by
$7\%$ for $w=0.2\pi/d$, and by $26\%$ for $w=2\pi/d$.

\begin{table}
\caption[t2]{Dependence on basis size of the calculated width $w_{\nu_{\rm
max}}$ of the swallow tail at the zone center. The table shows $w_6$ in
units of $\pi/d$, and the difference $w_{\nu_{\rm max}}/w_6 -1$ in per
cent.}
\label{tbl2}
\begin{tabular}{ll@{\ \ }|@{\ \ }c@{\ \ }|@{\ \ }c@{\ \ \ \ }c@{\ \ \ \
}c@{\ \ \ \ }c@{\ \ \ \ }c}
&&&\multicolumn{5}{c}{Difference} \\
&&&\multicolumn{5}{c}{for $\nu_{\rm max}$} \\
$V_0$ & $nU_0$ &$w_6$& 1 & 2 & 3 & 4 & 5\\
\hline
$E_0$ & $E_0$ & 0.077969 &  -0.95 & 4.36  &  0.05 & 0.00 & 0.00\\
$E_0$ & $4E_0$ & 0.373789 &  3.97 & 13.4  &  0.16 & 0.03 & 0.00\\
$E_0$ & $8E_0$ & 0.726953 &  10.5 & 24.0  &  0.36 & 0.17 & 0.01\\
$4E_0$ & $8E_0$ & 0.432695 &  0.86 & 34.0  &  0.97 & 0.87 & 0.06\\
$8E_0$ & $8E_0$ & 0.155234 &  -20.3 & 68.4  &  3.32 & 2.52 &
0.33\\
\end{tabular}
\end{table}

For the swallow tail at the zone boundary, Fig.~\ref{Width}, the analysis
is further simplified by taking $\phi=\pi/2$, which is an exact solution
at
$k=\pi/d$ \cite{bronski1,bronski2}. The value $\theta_0$ at which the
energy is stationary varies continuously around the loop.
The tip of the swallow tail is found by setting the derivative of $k$ with
respect to $\theta_0$ equal to 0. This yields
$\sin 2\theta_0=-(V_0/nU_0)^{1/3}$, and a width
\begin{eqnarray}
\label{widthformula}
w=\frac{\pi}{2d}\left[\left(\frac{nU_0}{E_0}\right)^{2/3}
-\left(\frac{V_0}{E_0}\right)^{2/3}\right]^{3/2}.
\end{eqnarray}
Compared to the numerical result in Fig.~\ref{Width} the contours
predicted by Eq.~(\ref{widthformula}) are shifted towards lower values of
$nU_0$, i.e., the width is
somewhat overestimated by Eq.~(\ref{widthformula}). For
$nU_0=2E_0$ and $V_0=E_0$ the width given by Eq.~(\ref{widthformula}) is
21\% above
the numerical result and, for
$nU_0=4E_0$ and $V_0=2E_0$, 38\% above. The model is thus most precise for
small widths.

\subsection{Physical understanding of swallow
tails}\label{soliton}

A striking feature of the results above is that the width of the swallow
tail
is nonzero as the strength of the periodic potential tends to zero.  As we
shall now describe, the states on the upper edge of the swallow tail then
correspond to periodic soliton solutions of the Gross-Pitaevskii equation
first discussed for a condensate in the absence of an external potential
by
Tsuzuki \cite{tsuzuki}.  For potentials having the form of Jacobi elliptic
functions, analytical results for periodic solitons have been obtained in
Ref.\ \cite{bronski2}.

The simplest case to think about is that at the zone
boundary, $k=\pi/d$.  The solution is then an equally spaced array of dark
solitons (that is solitons for which the wave function vanishes on some
surface), with one soliton per lattice spacing of the periodic potential.
That this solution has wave vector $\pi/d$ may be seen from the fact that
the phase difference across a dark soliton is $\pi$.  Since there is
one soliton per lattice spacing, the wave vector $k$, which is the average
phase
change
per unit length, is thus $\pi/d$.  For states
with this value of the wave vector, the energy of the highest state in the
first band and the energy of the state in the second band are degenerate
if
the lattice potential is absent, but the degeneracy is broken
when a weak periodic potential is applied.  The
state having dark solitons with centers at $x= \tau d$, where $\tau$ is an
integer, has a lower energy, since the solitons, which are density
rarefactions, are located at the maxima of the lattice potential.  On the
other hand, the state having solitons with centers at $x= (\tau+1/2) d$
has
a
higher energy.

The situation at $k=0$ may be described in similar terms.
Here the solutions on the upper edge of the swallow tail correspond to
periodic solitons with two dark solitons for every lattice period.  The
phase
change per lattice period is thus $2\pi$, which corresponds to a wave
vector
$k=0$ in the reduced zone scheme.  When the lattice potential is applied,
there is no change in the energy to first order in the lattice potential
because neighboring dark solitons are separated by $d/2$ and therefore the
energy due to the lattice potential vanishes, since the potential is
purely
sinusoidal.  However, there will be a contribution to second order in
$V_0$
since the lattice potential will make the spacings between neighboring
solitons unequal, even-numbered solitons being displaced in one direction,
and
odd-numbered ones in the other direction.  One may say that the lattice
potential causes a dimerization of the soliton array.  This picture gives
another way of understanding the conclusion arrived at earlier that the
splitting between a state on the upper edge of the swallow tail and that
in
the next higher band is proportional to $V_0$ for $k=\pi/d$ and
proportional
to $V_0^2$ for $k=0$.

Let us now consider states with $k \neq 0$, $\pi/d$.  In the absence of a
lattice potential, there exist solutions of the Gross-Pitaevskii equation
which are periodic arrays of gray solitons.  For these, the density never
vanishes, and the phase change across the soliton is less than $\pi$.
With the boundary conditions that are usually imposed, such solitons
move with a uniform velocity, which we denote by
$v_{\rm soliton}$.  However, by boosting the velocity of the condensate by
a
constant value $-v_{\rm soliton}$ everywhere, the solution becomes a
stationary one,
and the positions of the solitons
will remain fixed. The lattice potential
will then lift the degeneracy of the energy with respect to translation of
the
periodic soliton, just as it did in the earlier example of an array of
dark
solitons.  The wave vector of the condensate is then obtained by combining
the
phase differences due to the solitons with the spatially-dependent phase
due
to the velocity shift, $-v_{\rm soliton}$.  As the wave vector departs
increasingly from
$k=0$ or $\pi/d$, the velocity $v_{\rm soliton}$ and the minimum density
in the soliton
increase.  Eventually, the density modulation in the soliton drops to
zero,
and the periodic soliton branch merges with that for motion of a uniform
condensate.  If the coherence length $\xi =
\hbar/\sqrt{2mnU_0}$ is much less than the lattice spacing, that is $nU_0
\gg
E_0$, solitons are generally well separated, and the highest value of
$v_{\rm
soliton}$ is equal to the sound velocity, $(nU_0/m)^{1/2}$.

The physical picture of states corresponding to the swallow tail gives
insight
into the convergence of the wave functions as the size of the basis is
increased.  The characteristic dimension of an isolated soliton is of
order
three times the coherence length, and therefore
the
ratio of the width of the soliton compared with the lattice spacing is of
order
$(E_0/nU_0)^{1/2}$.  Thus to give a good account of the structure of a
soliton,
one would expect to need $\nu_{\rm max}\sim (nU_0/E_0)^{1/2}$.  For the
parameter
values that we have studied, this is consistent with the fact that the
numerical calculations were well converged for $\nu_{\rm max}=3$.

\section{Elementary excitations and stability}
\label{SecElex}
In the preceding section we have explored the nature of Bloch
waves throughout the Brillouin zone, for different values of the
mean-field interaction and the lattice potential. In
the present
section we investigate their stability
to small perturbations. We shall treat both energetic
and dynamical instabilities, starting from the Gross-Pitaevskii
energy functional. Subsequently, we shall examine the nature of the
excitations at long wavelengths,  and we shall derive stability
criteria from hydrodynamic equations.

\subsection{Energetic stability}
To investigate energetic stability of Bloch states, we expand the
Gross-Pitaevskii energy functional (\ref{energy})
to second order in the deviation $\delta\psi$ of the condensate wave
function from the equilibrium  solution $\psi_0$, subject to the
condition that the total particle number be fixed. To satisfy
the constraint, it is convenient to work with the
thermodynamic potential $G=E-\mu N$, where $N$ is the particle
number, and to allow the variations of $\psi$ to be arbitrary. Writing
$\psi=\psi_0+\delta\psi$ and expanding $G$ to second order in $\delta
\psi$, one finds
\begin{equation}
G=G[\psi_0] +\delta G_1+\delta G_2.
\end{equation}
The first-order term vanishes when $\psi_0$ satisfies
the time-independent Gross-Pitaevskii equation (\ref{gp}). The
second-order term is
\begin{eqnarray}\label{fluc}
&&\delta G_2=\nonumber\\
&&\int{d{\bf r}
\left(\delta\psi^*[-\frac{\hbar^2}{2m}\nabla^2+V ({\bf
r})-\mu]\delta\psi\right.}
\\
&&\left. +\frac{1}{2}U_0\left[(\psi_0^*)^2(\delta\psi)^2+
      \psi_0^2(\delta\psi^*)^2
+4|\psi_0|^2\delta\psi\delta\psi^*\right]\right).\nonumber
\end{eqnarray}
This quadratic form can be written in a compact matrix notation as
\begin{equation}
\delta
G_2=\frac{1}{2}\int{d{\bf r}\delta\Psi^{\dagger}\hat{A}\delta\Psi},
\label{second}
\end{equation}
where we have introduced the column vector
\begin{equation}
\delta\Psi= \left(\begin{array}{c}
      \delta\psi\\
\delta\psi^*
\end{array}\right)
\end{equation}
and the matrix
\begin{equation}
\hat{A} = \left(\begin{array}{cc}
      L&U_0\psi_0^2\\
U_0(\psi_0^*)^2&L
\end{array}\right),
\label{A}
\end{equation}
which is Hermitian.
The operator $L$ occurring in (\ref{A}) is given by
\begin{equation}
L=T+V+2U_0|\psi_0|^2-\mu \label{L},
\end{equation}
with $T=-\hbar^2\nabla^2/2m$ being the kinetic
energy operator.

The solutions to the Gross-Pitaevskii equation
correspond to stationary values of the thermodynamic potential. In
order to investigate the stability of these solutions under small
perturbations we must look at the second order term
(\ref{second}). When this term is positive for all $\delta\psi$
the solution is energetically stable.
The system is stable if the equation
\begin{equation}
\hat{A}\delta\Psi=\lambda\delta\Psi \label{eigen}
\end{equation}
has only positive eigenvalues. Instability sets in when the lowest
eigenvalue vanishes.

\subsection{Dynamical stability}
\label{dynstab}

To explore dynamical instability, we need to examine eigenvalues
of the time-dependent Gross-Pitaevskii equation,
\begin{equation}
-\frac{\hbar^2}{2m}\nabla^2\psi+V({\bf r})\psi +U_0|\psi|^2\psi
=i\hbar\frac{\partial\psi}{\partial t},
\label{gpt}
\end{equation}
and its complex conjugate.
The pair of equations obtained by making the substitution
$\psi=\psi_0+\delta
\psi$ in this equation and its complex conjugate may, when linearized,
be written in the matrix form
\begin{equation}
i\hbar\frac{\partial\delta\Psi}{\partial
t}=\hat{\sigma}_z\hat{A}\delta\Psi \label{timedep}
\end{equation}
with $\hat{\sigma}_z$ being the Pauli matrix in the usual representation,
\begin{equation}
\hat{\sigma}_z=
\left(
\begin{array}{cc}
1 & 0 \\ 0 & -1
\end{array}
\right) . 
\end{equation}
Note that
$\hat{\sigma}_z\hat{A}$ is non-Hermitian, and therefore can have both real
and
complex eigenvalues. Complex eigenvalues always occur in pairs, since if
$\delta\Psi_\lambda$ is an eigenfunction of
$\hat{\sigma}_z\hat{A}$ with eigenvalue $\lambda$, then
$\hat{\sigma}_z\delta\Psi_\lambda^*$ is an eigenfunction with eigenvalue
$\lambda^*$. Thus if the matrix has complex eigenvalues, there is always
one
eigenvalue with a positive imaginary part, and therefore
the corresponding mode will
grow exponentially in time. The system is then dynamically unstable.

\subsection{Hydrodynamic analysis}\label{LWanalysis}

To study excitations which have wavelengths much greater than the
lattice spacing, it is possible to use a hydrodynamic approach.
One works with the average particle density ${\bar
n}({\bf r})$ and an average phase to be defined below, where the
averages are to be taken over a volume having linear dimensions much
greater
than the lattice spacing but still much smaller than the wavelength
of the disturbance.  Such an approach has previously been employed for
small
condensate velocities in Ref.~\cite{kraemer}.

Consider a condensate subjected to a potential which consists of the sum
of
two
contributions, one due to
the lattice potential and another one which varies slowly in space on the
scale of the
spacing of the optical lattice. We denote the latter contribution by
$\bar{V}$. If $\bar{V}$ is spatially uniform, the phase $\phi({\bf r}, t)$
of
the condensate wave function
in a stationary state may be written as the sum of a spatially-varying
part,
$\phi_0({\bf r})$ and a time-dependent part $\chi(t)$,
\begin{equation}
\phi({\bf r}, t)= \phi_0({\bf r}) +  \chi(t).
\label{phase}
\end{equation}
Observe that in writing the phase in this form we have nowhere made any
assumption about
how fast $\phi_0$ changes over distances of the order of the lattice
spacing.
The phase evolves in
time according to the Josephson equation
\begin{equation} \hbar\frac{\partial{ \phi}}{\partial t}
=\hbar\frac{\partial{\chi}}{\partial t}
=-\mu -\bar{V},
\label{josephson}
\end{equation}
where $\mu$ is the chemical potential calculated for $\bar{V}=0$, that is
with
only the lattice potential acting \footnote{We here define the chemical
potential by the
equation $\mu(n, k_x)=\partial {\tilde E}/\partial n$.  With this
definition, the chemical potential depends on the wave vector  of
the condensate. For a Galilean-invariant system, $\mu (n, k_x)= \mu
(n, 0) +m\hbar^2 k_x^2/2m$.  Our definition of the chemical
potential is therefore different from the one conventionally used
for Galilean-invariant systems, where the chemical potential is
usually defined in terms of the energy of the system in the frame
in which it is locally at rest.}.

When the average particle density and the potential $\bar{V}$ vary slowly
in
space
on length scales large compared with
the lattice spacing, one expects the time rate of change of the phase
$\chi$ to
be
given by the same result, except that the chemical potential and $\bar{V}$
now both
vary in space.  In the presence of inhomogeneity, the phase will evolve
with
time at
different rates at different points in space, thereby ``winding up" the
phase
difference between different spatial points. The wave vector of the
condensate wave function is determined by the average rate at which the
phase
of the
wave function advances in space.
  Thus
the change $\delta \bar{\bf k}({\bf r})$ in the wave vector of the
condensate
is given by
\begin{equation}
\delta\bar{\bf k}({\bf r}) =
\boldsymbol{\nabla}
\chi ({\bf r}).
\label{wavenumber}
\end{equation}
It therefore follows from Eq.\ (\ref{josephson}) that the equation for
the rate of change of the
wave vector has the form
\begin{equation}
\hbar\frac{\partial \bar{\bf k}}{\partial t}
=-\boldsymbol{\nabla}[\mu({ n},{\bf k}) +\bar{V}({\bf r}, t)]    .
\label{dkdt}
\end{equation}
When spatial variations are slow, it is a good approximation to
assume that the energy density locally is given by the expression
for the energy density of a uniform system, but with
spatially-varying local densities and wave vectors:
\begin{equation}
E=\int d{\bf r}\left[{\tilde E}({\bar n}({\bf r}),\bar{{\bf
k}}({\bf r}))+{\bar n}({\bf r})\bar{V}\right]. \label{energyhydro}
\end{equation}
Here ${\tilde E}(n,{\bf k})$ is the energy density of the
state of the uniform system having a wave vector
$\bf k$ and particle density $n$ when $\bar{V}=0$. In this approximation,
the
chemical
potential is that of a bulk system having a wave vector and average
density
equal to the values locally in the nonuniform system. To simplify the
notation
we
omit the bars in the following, but it should be remembered that the
symbols $n$ and
$\bf k$ always refer to average values locally. At the same level of
approximation, the local
current density, the flux of particle number per unit area, is given
by the result for a uniform system,
\begin{equation}
{\bf j}= \frac{1}{\hbar}{\boldsymbol\nabla_{\bf k}}{\tilde
E}({n},{\bf k}).
\label{current}
\end{equation}
Thus the equation of continuity is
\begin{equation}
\frac{{\partial}{ n}}{\partial t}
+\boldsymbol{\nabla}\cdot{\bf j} =\frac{{\partial}{
n}}{\partial t}+\frac{1}{\hbar}\boldsymbol{\nabla}\cdot
\boldsymbol{\nabla}_{\bf k}{\tilde E}({ n},{\bf k}) =0.
\label{continuity}
\end{equation}
To find the elementary excitations, we now linearize Eqs.\ (\ref{dkdt})
and
(\ref{continuity}).
We denote changes in the local density by $\delta { n}$, those in the wave
vector by $\delta {\bf k}$, and those in the potential by $\delta
\bar{V}$.
If one looks for solutions varying in space and time as $\exp {i({\bf
q}\cdot
{\bf r}-\omega t)}$, one finds that
\begin{equation} ({\tilde E}_{n,
{\bf k}}\cdot{\bf q} -\hbar \omega)\delta { n} + {\bf
q}\cdot{\tilde E}_{{\bf k},{\bf k}}\cdot \delta { {\bf k}}=0
\,\,\,\,\,\,\,\,\,\,\,\,\,
\label{lin1}
\end{equation}
and
\begin{equation}
{\tilde E}_{n,n}{\bf q}\delta{ n} + {\bf
q}{\tilde E}_{n,{\bf k}}\cdot \delta { {\bf k}} -\hbar \omega
\delta { {\bf k}}=-{\bf q}\delta\bar{V}.
\label{lin2}
\end{equation}
Here
\begin{equation}
{\tilde E}_{n,n}=\frac{\partial^2 {\tilde E}}{\partial n^2}=\frac{\partial
\mu}{\partial n}.
\end{equation}
The derivative
\begin{equation}
{\tilde E}_{k_i,k_j}=\frac{\partial^2 {\tilde E}}{\partial
k_i\partial k_j} \equiv n \hbar^2\left(\frac{1}{m}\right)_{ij}
\end{equation}
is, apart from factors, a generalization of the usual effective mass
tensor for a single particle.  Note that it depends on the particle
density and on the wave vector of the superfluid flow.
The final derivatives are
\begin{equation}
{\tilde E}_{n,k_i}=\frac{\partial^2 {\tilde E}}{\partial
n\partial k_i} = \frac{\partial \mu}{\partial
k_i}=\hbar\frac{\partial j_i}{\partial n}.
\label{derivativeEnk}
\end{equation}
In the absence of the lattice potential,
it follows from Galilean invariance that the contribution to the energy
per particle that depends on
the wave number is given by $\hbar^2 k^2/2m$.  Consequently,
the derivative (\ref{derivativeEnk}) reduces to the
condensate velocity (times $\hbar$).
All
derivatives
are to be evaluated for the unperturbed value of the density $ n$
and of the wave vector ${\bf k}$.

The above discussion applies for arbitrary directions of $\bf k$
and $\bf q$.  Let us now apply the results to the case where both
these vectors are in the $x$ direction.  The eigenfrequencies of the
system
are found by solving Eqs.\ (\ref{lin1}) and  (\ref{lin2}) with $\delta
\bar{V}=0$, and they are given by
\begin{equation}
\hbar \omega = q_x{\tilde E}_{n,k}
\pm ({\tilde E}_{n,n} {\tilde E}_{k,k}q_x^2)^{1/2}.
\label{critvel}
\end{equation}
Equation (\ref{critvel}) provides a generalization to current-carrying
states
of results derived in Ref.~\cite{kraemer} for a condensate initially at
rest.
In order to elucidate the meaning of Eq.\ (\ref{critvel}), let us first
consider the case of $k \rightarrow 0$.  The energy per particle is
quadratic
for small $k$, and therefore ${\tilde E}_{n,k}$ tends to zero in this limit
and
\begin{equation}
\hbar \omega =
\pm ({\tilde E}_{n,n} {\tilde E}_{k,k}q_x^2)^{1/2}.
\label{latticesound}
\end{equation}
The modes are then sound waves, with the sound speed given by
\begin{equation}
s=
({\tilde E}_{n,n} {\tilde E}_{k,k})^{1/2},
\label{soundvelocity}
\end{equation}
where the derivatives are to be evaluated for $k \rightarrow 0$.
This result agrees with that obtained in Ref.\ \cite{kraemer}.
For a translationally invariant system,
${\tilde E}(n, k)=\mbox{${\tilde E}(n,k=0)$}+n\hbar^2 k^2/2m$.  Therefore
$ {\tilde E}_{n,n} ={\partial \mu}/{\partial n} =U_0$ and ${\tilde
E}_{k,k}=n\hbar^2/m$, and the sound velocity is thus given by the usual
result for a homogeneous gas
\begin{equation}
s=\sqrt{\frac{nU_0}{m}}.
\label{soundveluni}
\end{equation}
The mixed derivative ${\tilde
E}_{n,k}=\hbar^2k/m =\hbar v$,
where $v=\hbar k/m$ is the velocity of the fluid.
The expression~(\ref{critvel}) for the frequency then reduces to the
familiar result
\cite{landau}
\begin{equation}
\hbar \omega = {\bf q \cdot v}
\pm sq.
\end{equation}
Observe that for a condensate moving in
an
optical lattice, the quantity ${\tilde E}_{n,k}/\hbar=\hbar^{-1} \partial
\mu/\partial k =\partial j/\partial n$ takes the place of the mean flow
velocity $j/n$ that occurs in the analogous result for a translationally
invariant system.

Let us now consider the stability of the system to long-wavelength
perturbations of the local density and wave vector.  The system is
energetically unstable if such perturbations can lead to a reduction of
the
energy. In the absence of the potential, the functional for the  energy
may
be
expanded about the original state, and one finds
\begin{eqnarray}
E=E_0 + \int d{\bf r} \{\mu\delta n +  \hbar j \delta k  \nonumber\\+
\frac{1}{2} \left[{\tilde E}_{n,n}(\delta n)^2+ 2{\tilde
E}_{n,k} \delta n \delta k+ {\tilde E}_{k,k}(\delta k)^2
\right]\}.
\end{eqnarray}
The first order terms vanish if the total number of particles and the
phase
of the wave function at the boundaries are fixed. The latter condition
implies that the change in the total particle current is unaltered.
The conditions for the quadratic form to be positive definite are that
\begin{eqnarray}
{\tilde E}_{n,n}>0,\;\;
{\tilde E}_{k,k}>0
\label{stab1}
\end{eqnarray}
and
\begin{eqnarray}
{\tilde E}_{n,n}
{\tilde E}_{k,k}> ({\tilde E}_{n,k})^2.
\label{stab2}
\end{eqnarray}
Sufficient conditions for energetic stability are that the condition
(\ref{stab2}) and one of the
conditions (\ref{stab1}) are satisfied, since the other inequality is then
satisfied automatically.  Observe that when condition (\ref{stab2})
becomes
an
equality, the system has a zero-frequency mode.

The numerical calculations to be described in the next section
indicate that energetic
instability sets in first at long wavelengths ($q\rightarrow0$).
Consequently,
Eqs.~(\ref{stab1}) and (\ref{stab2}) are the general conditions for
energetic
stability.
As an example, we shall use the condition (\ref{stab2}) in
Sec.~\ref{Stab1D} below
to determine an approximate criterion for the limit of stability at the
zone
boundary.

The condition for onset of dynamical instability is that
the eigenfrequency~(\ref{critvel}) become complex, which occurs if
either ${\tilde
E}_{n,n}$ or ${\tilde E}_{k,k}$ become negative.
The first condition corresponds to the compressibility being negative,
the second to the
effective mass being negative.

\section{Stability of Bloch waves in a one-dimensional lattice}
\label{Stab1D}

The stability considerations in the previous section were general
in nature. In the following we apply them to a one-dimensional
potential $V(x)=V_0\cos(2\pi x/d)$ with period $d$. As in Ref.~\cite{WDN}
we consider
changes in the condensate wave function of the form
\begin{equation} \delta\psi= {\rm e}^{ikx}[u_q(x){\rm e}^{iqx}+
v^*_q(x){\rm e}^{-iqx}],\label{fluctu}
\end{equation} where $u_q(x)$ and $v_q(x)$ have the periodicity of
the lattice. As a result, Eq.\ (\ref{second}) becomes
\begin{equation}
\delta
G_2=\frac{1}{2}\int{d{\bf r}\delta\Phi^{\dagger}\hat{B}\delta\Phi},
\label{second1D}
\end{equation}
where
\begin{equation}
\delta\Phi= \left(\begin{array}{c}
      u_q\\
v_q
\end{array}\right)
\end{equation}
and
\begin{eqnarray}
\label{B}
\hat{B} =
\left(\begin{array}{cc}
      L_+&U_0f_0^2\\
U_0(f_0^*)^2&L_-
\end{array}\right),
\end{eqnarray}
with $f_0(x)=\exp(-i k x)\psi_0(x)$ (cf. Eq.~(\ref{Blochwave})). The
operators $L_{\pm}$ is given by
\begin{equation}
L_{\pm}= -\frac{\hbar^2}{2m}\left(\frac{d}{dx}+i(\pm
k+q)\right)^2+ V(x)-\mu+2U_0|\psi_0|^2.
\label{pm}
\end{equation}

According to Eq.~(\ref{timedep}) the linearized time-dependent
Gross-Pitaevskii equation becomes
\begin{equation}
i\hbar\frac{\partial\delta\Phi}{\partial
t}=\hat{\sigma}_z\hat{B}\delta\Phi.
\end{equation}

As discussed above, the stability of solutions
to the time-independent Gross-Pitaevskii
equation may be determined by study of
the eigenvalues of the operators $\hat{B}$ and $\hat{\sigma}_z\hat{B}$.
Energetic
instability sets in when $\hat{B}$
first acquires a zero eigenvalue, while dynamical instability sets in when
one
of the eigenvalues of $\hat{\sigma}_z\hat{B}$ becomes complex.
Before presenting numerical results, we give two analytical examples of
the
use of the method. First, we consider the case
$V(x)=0$, and then we derive an approximate condition for stability of
states
at the zone boundary.

\subsection{The homogeneous Bose gas}

The problem of instability of a homogeneous Bose gas has previously been
considered in Ref.~\cite{WN2}. The calculations described in this
subsection are similar to those of Ref.~\cite{WN2}, but we offer a
somewhat different physical interpretation.

For a homogeneous gas $(V(x)=0)$,  the solutions to the Gross-Pitaevskii
equation take the form
\begin{equation}
\psi_0=\sqrt{n}{\rm }{\rm e}^{ikx},
\end{equation}
where $n$ is the density. The corresponding chemical potential is
$\mu=nU_0+\hbar^2k^2/2m$. The change $\delta\psi$ in the condensate wave
function is written
in the form (\ref{fluctu}), and since the system is uniform we
look for solutions $u_q$ and $v_q$ that are constant in space. The
 matrix $\hat{B}$ given by Eq.~(\ref{B}) then becomes
\begin{eqnarray}
\label{B1}
&&\hat{B} =\nonumber\\
&&\left(\begin{array}{cc}
      nU_0+\frac{\hbar^2}{2m}(q^2+2kq)&nU_0\\
      nU_0&nU_0+\frac{\hbar^2}{2m}(q^2-2kq)
\end{array}\right).\nonumber\\
&&
\label{Bhomegeneous}
\end{eqnarray}
The stability limit is obtained from the
condition that the determinant of the matrix vanish, corresponding
to the existence of a zero eigenvalue. This yields
\begin{equation}
      (\hbar^2kq/m)^2 =(\hbar^2q^2/2m
+nU_0)^2-(nU_0)^2 \equiv \epsilon_{q}^2,\label{landau}
\end{equation}
where $\epsilon_{q}$ is the Bogoliubov result for the energy of an
excitation in a dilute Bose gas.  Thus the condition is equivalent
to the Landau criterion that the minimum velocity at which it is
energetically favorable to create excitations is given by
$\epsilon(q)/\hbar q$. On division by $q^2$ the condition
(\ref{landau}) becomes
\begin{equation}
k^2=q^2/4 +(ms/\hbar)^2,
\label{landau1}
\end{equation}
where the sound
velocity $s$ is given by
Eq.\ (\ref{soundveluni}).

In addition to the energetic instability considered
above the system may develop a dynamical instability.
The dynamical instability exists only when the
periodic potential is present, since otherwise there is no mechanism for
transferring momentum to the fluid.

In order to understand the origin of the dynamical instability for a weak
periodic potential let us first consider the eigenvalues of
$\hat{\sigma}_z\hat{B}$ in the absence of a periodic potential. In this
case the matrix
$\hat{\sigma}_z\hat{B}$ is obtained from $\hat{B}$ by
changing the sign of the matrix elements in the second row of
Eq.~(\ref{Bhomegeneous}). Its eigenvalues $\lambda$ are given by
\begin{equation}
\lambda_{\pm}=\frac{\hbar^2kq}{m}\pm\left(nU_0\frac{\hbar^2q^2}{m}
+\frac{\hbar^4q^4}{4m^2}\right)^{1/2} .\label{eigenval}
\end{equation}
As usual, the physical excitation energies of the system correspond to
the plus sign in this equation.  Note that this expression becomes
identical with Eq.\ (\ref{critvel}) in the long-wavelength limit
($q\rightarrow
0$).
The eigenvalues (\ref{eigenval}), which are obtained for the case
when the periodic potential is absent, are always real when $U_0$ is
positive. Thus, as remarked above, for repulsive interactions there is no
dynamical
instability in the absence of a periodic potential.

Now, let us consider the case of a weak periodic potential.
The appearance of a complex
eigenvalue corresponds to a resonance in which two
phonons are created. The resonance condition requires the total momentum
of the phonons to be $G=2\pi/d$ (or $-G$) while their total energy must be
zero. This implies that
\begin{equation}
\lambda_+(q)+\lambda_+(G-q)=0
\end{equation}
or
\begin{equation}
|k|=\frac{1}{G}\left(\sqrt{k_0^2q^2+q^4/4}+\sqrt{k_0^2(G-q)^2+(G-q)^4/4}\right),
\label{wavenum}
\end{equation}
where $k_0=ms/\hbar$. The momenta of the two phonons are opposite that
of the flow.
The condition (\ref{wavenum}) is precisely the Landau condition
for the creation of a pair of excitations with total wave number $G$ in a
superfluid flowing with velocity $\hbar k/m$. According to Eq.\
(\ref{wavenum}), the magnitude of the wave number,
$|k|$, for resonance decreases  from $(k_0^2+G^2/4)^{1/2}$ to
$(k_0^2+G^2/16)^{1/2}$ as $q$ increases from 0 to $G/2$.  With further
increase in $q$, the
 magnitude of the resonant wave number increases, since it
is symmetric with respect to the interchange of $q$ and $G-q$.
When
the lattice potential is weak,
dynamical instability therefore first appears when $q=G/2 $ at
a wave number $k$ given by $|k|=(k_0^2+G^2/16)^{1/2}$.

In the next section we show how the thresholds for energetic and
dynamical instability are calculated for a non-vanishing periodic
potential for different values of the wave number $k$ as functions of
$nU_0$ and $V_0$.
The above prediction for the onset of the dynamical instability for
$V_0\rightarrow 0$,
$|k|=(k_0^2+G^2/16)^{1/2}=(\pi/d)(nU_0/2E_0+1/4)^{1/2}$, is an exact
result, first derived in Ref.~\cite{WN2}, and may be compared to the
asymptotic value of the numerical results (solid lines) in
Fig.~\ref{DynamicStab} in the limit of
$V_0 \rightarrow 0$. The numerical results agree with the analytical
expression within the precision of the calculation.

\subsection{Stability of states at the zone boundary}
A second illustrative example is to consider the condition for
long-wavelength instabilities to arise in a flow for which $k=\pi/d$.
This may be done using the hydrodynamic formalism described in
Sec.~\ref{LWanalysis}.  As shown in Refs.~\cite{bronski1} and
\cite{bronski2}, for $k=\pi/d$ there is an exact solution to the
time-independent Gross-Pitaevskii equation of the form
\begin{equation}
\psi =\sqrt{n}{\rm e}^{i k x} (\cos \theta  +\sin \theta {\rm e}^{-i
2\pi x/d}),
\label{exactsolution}
\end{equation}
which is the same as Eq.\ (\ref{trial}) with $\phi=\pi/2$.
To investigate the stability of the state with $k=\pi/d$, we require the
solution also for $k$ in
the vicinity of $\pi/d$ in order to evaluate the derivatives with respect
to
$k$, and we shall assume that this is given by Eq.\ (\ref{exactsolution}),
with $\theta$ being treated as a variational parameter.

We start from the energy function (\ref{en1}), and for the trial
function
(\ref{exactsolution}) the kinetic energy per particle is given by
\begin{equation} \epsilon_{\rm kin}= 4E_0 \left(\kappa^2
+\sin^2\theta-2\kappa \sin^2\theta\right), \label{en22}
\end{equation}
where $\kappa=kd/2\pi$. The potential energy per particle is
\begin{equation}
\epsilon_{\rm pot}=V_0\sin\theta\cos\theta,
    \label{en33}
\end{equation}
while the interaction energy is
\begin{equation}
\epsilon_{\rm
int}=\frac{nU_0}{2}\left(1+2\cos^2\theta\sin^2\theta\right).
\label{en44}
\end{equation}
 The energy per particle is stationary when
\begin{equation}
(\kappa-\frac{1}{2})\sin 2\theta=\frac{V_0}{8E_0}\cos 2\theta+
\frac{nU_0}{8E_0}\cos2\theta\sin
2\theta. \label{iter}
\end{equation}
For $k=\pi/d$, the solution of Eq.\ (\ref{iter}) is $\sin
2\theta=-V_0/nU_0$.
On inserting this result into Eqs.\ (\ref{en22}-\ref{en44}), we obtain
the total energy per particle ${\tilde E}/n$, and
for $k=\pi/d$ we get
\begin{equation}
{\tilde E}=nE_0+\frac{n^2}{2}U_0-\frac{V_0^2}{4U_0}
\end{equation}
and
\begin{equation}
\frac{\partial^2{\tilde E}}{\partial n^2}=U_0. \label{part1}
\end{equation}
The other two derivatives are most easily evaluated by using the fact that
$\partial {\tilde E}/\partial k$ is $\hbar$ times the particle current
density
(see Eq.\ (\ref{current})).  The current may be calculated directly from
the
trial wave function.  For definiteness we consider states with positive
$k$,
and the result is
\begin{equation}
j=\frac{n\hbar}{m}\left(k -\frac{2\pi}{d} \sin^2 \theta\right).
\label{current2}
\end{equation}
The derivatives of $\theta$ with respect to $n$ and with respect to $k$
may
be calculated by differentiating Eq.\ (\ref{iter}), and one finds
\begin{equation}
\frac{\partial^2{\tilde E}}{\partial n\partial
k}=\frac{\hbar^2\pi}{md}\frac{nU_0}{[(nU_0)^2-V_0^2]^{1/2}} \label{part2}
\end{equation}
and
\begin{equation}
\frac{\partial^2{\tilde E}}{\partial
k^2}=\frac{n\hbar^2}{m}\left(1-\frac{4E_0V_0^2}{nU_0[(nU_0)^2-V_0^2]}
\right).
\label{part3}
\end{equation}
These derivatives are then inserted into the condition
(\ref{stab2}), and the boundary of the region of stability
is given by
\begin{eqnarray}\label{AMstablimit}
V_0^2=\left(nU_0\right)^2\frac{nU_0-2E_0}
{nU_0+4E_0}.
\end{eqnarray}
As we shall see below, this curve deviates by no more than 9.5\% from the
one calculated numerically, i.e., the contour for $k=\pi/d$ in
Fig.~\ref{LandauStab}.

\subsection{Numerical calculations of stability limits}

We now describe results of a
stability analysis of the stationary state solutions
$\psi_0$ found in Sec.~\ref{SecBW}. The amplitudes $u_q$ and $v_q$
in Eq.~(\ref{fluctu}) are expanded in terms of plane waves:
\begin{eqnarray}
u_q=\sum^{ l_{\rm max}}_{ l=- l_{\rm max}}
     {u_{ l,q}{\rm e}^{i 2\pi l x/d}} \label{StabExpansion}
\end{eqnarray}
and
\begin{eqnarray}
v_q=\sum^{ l_{\rm max}}_{ l=- l_{\rm max}}
     {v_{ l,q}{\rm e}^{i 2\pi l x/d}} .
\end{eqnarray}
In the sums, we take $l_{\rm max}$ to be less than $ \nu_{\rm max}$.  If
this is not done,
spurious instabilities can result.  These merely express the fact that the
condensate wave function has not been optimized for modes with wavelengths
less than $d/\nu_{\rm max}$.
The operator
$\hat{B}$ in Eq.~(\ref{B}) is represented as a matrix of dimension
$4 l_{\rm max}+2$ in terms of the plane-wave basis.

We now investigate the stability of states corresponding to points on the
swallow tail
in the lowest band. In the reduced zone
representation
used in Figs.~\ref{Bandstructure2}-\ref{Bandstructure3}, this swallow tail
is split into two or more pieces. However, by choosing the first Brillouin
zone as $0 \le k \le 2\pi/d$, this swallow tail will appear in
one piece for $w<2\pi/d$. In the discussion below, we shall use the latter
representation, and all values of $k$ will be taken to lie in the interval
$[0,2\pi/d]$. We shall not consider swallow tails that have a width
greater than $2\pi/d$.

\subsubsection{Energetic stability}

At the boundary of the region of stability, the operator $\hat{B}$
has a
vanishing eigenvalue and therefore its determinant vanishes. We evaluate
the conditions under which the determinant of $\hat{B}$ first vanishes by
using the  Mathematica$^{\circledR}$ routine ``Det". The calculation of
the
energetic
stability limits
proceeds as follows: we choose
values of $k$ and $V_0$, and determine the value of
$nU_0$ at which the determinant of $\hat{B}$ first becomes positive
definite
for all $q$. We find that
energetic instability occurs first for $q \rightarrow 0$, in agreement
with Fig.~1 of Ref.~\cite{WN2}.

\begin{figure}[t]
\includegraphics[width=80mm]{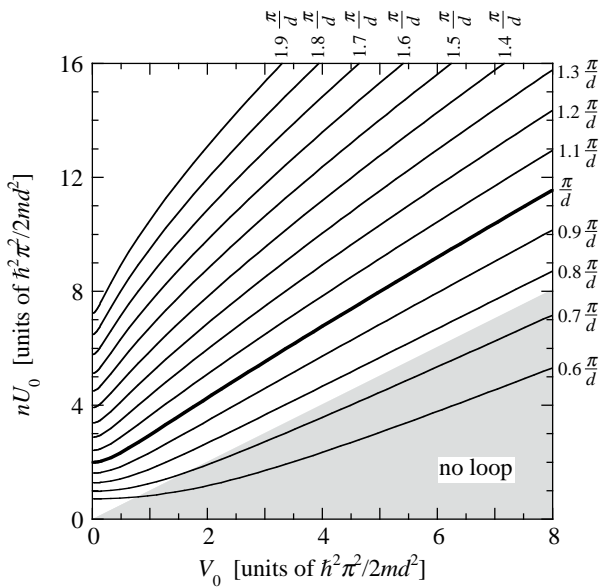}
\caption{\label{LandauStab}
Contour plot of the maximum wave vector $k$ for energetic stability.
Contours for $k\le \pi/d$ corresponds to states on the lowest branch of
the lowest band, and those for $k>\pi/d$ to states on the lower edge of
the swallow tail at the zone boundary.}
\end{figure}

Figure~\ref{LandauStab} is a contour plot of the maximum wave vector for
energetic stability as a function of $V_0$ and $nU_0$. Values of $k$ less
than $\pi/d$ correspond to the lowest energy states for a given $k$, while
higher values correspond to states on the lower edge of the swallow tail.
As the plot shows, the range of $k$ values for which states are stable
increases with increasing $nU_0$ and decreasing $V_0$. The numerical
calculations in Fig.~\ref{LandauStab} are converged to
within the thickness of the contour lines for $\nu_{\rm max}=3$ and $
l_{\rm
max}=3$.

A number of insights into the behavior of the contours may be obtained
from
analytical arguments.  First, the intercepts on the $nU_0$ axis of the
contours for the wave vector at which energetic stability sets in may be
determined from the Landau criterion.  For an interacting Bose gas with no
lattice, energetic stability occurs when the velocity of the gas becomes
equal to the sound speed $s$, Eq.\ (\ref{soundveluni}). The velocity of
the
gas is $\hbar k/m$, and therefore the condition is
\begin{equation}
\frac{\hbar k}{m} =s \;\;{\rm or}\;\;
\frac{nU_0}{E_0}=2\left(\frac{d}{\pi}\right)^2 k^2.
\end{equation}
The numerical results agree with this.

\begin{figure}[t]
\includegraphics[width=80mm]{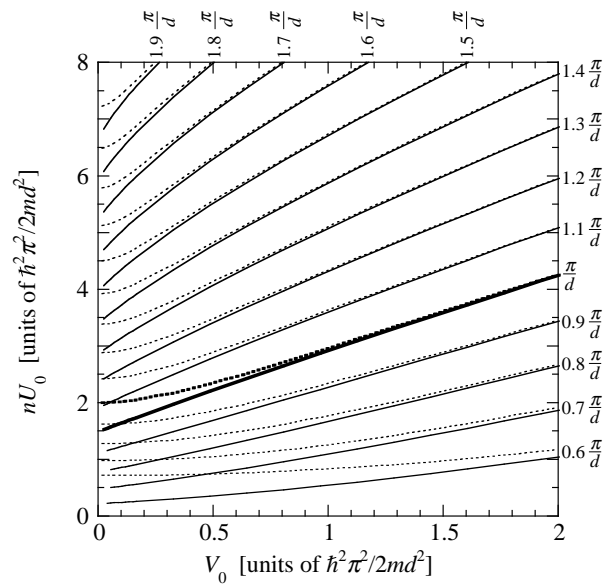}
\caption{\label{DynamicStab}
Contour plot of the maximum wave vector $k$ for dynamical stability (solid
lines). The notation is the
same as in Fig.\ \ref{LandauStab}. For comparison the corresponding
results for energetic stability are shown by dotted lines.}
\end{figure}

 A second general remark is that, for small $V_0$, one would expect on the
basis of perturbation theory  that the contours for energetic stability
would
behave as $V_0^2$, again in agreement with the numerical results.  
However,
for high values of $nU_0$, the quadratic dependence holds only for a
limited range of $V_0$. The contours are approximately linear at higher
values of $V_0$, just as are those for the width of the swallow tail at
the
zone boundary (see Fig.\ \ref{Width}).  The wave vector of the tip of the
swallow tail sets a natural limit to the  wave vector at which instability
sets in, and indeed this limit is approached for large $nU_0$.

A comparison of the width of the swallow tail (Fig.~\ref{Width}) and
the stability boundary (Fig.~\ref{LandauStab}) allow us to conclude
that, within the range of parameters investigated, states on the lower
edge
of the swallow tail at the
zone boundary are never stable for all $k$.
For the conditions under which the spectrum is given by
Fig.~\ref{Bandstructure2}a, instability sets in around
$k=1.7\pi/d$ and for the conditions appropriate for
the spectrum shown in Fig.~\ref{Bandstructure2}b, around
$k=1.1\pi/d$.

\subsubsection{Dynamical stability}

The boundary for dynamical stability is determined following a numerical
procedure similar to the one above for energetic stability.
On one side of the boundary, the operator $\hat{\sigma}_z \hat{B}$ has 
only
real eigenvalues
for all $q$, while on the other side it has some eigenvalues that are
complex. The eigenvalues of $\hat{\sigma}_z \hat{B}$ are determined using
the
Mathematica$^{\circledR}$ routine ``Eigenvalues".
With increasing $k$ (at fixed $nU_0$ and $V_0$), energetic instability
first occurs for $q\rightarrow 0$, while dynamical instability first sets
in at $q=\pi/d$. 
On the lowest branch of the lowest band, dynamical instability exists only
for $k>\pi/2d$, in agreement with Fig.~1 of Ref.~\cite{WN2}.

Figure~\ref{DynamicStab} shows a contour plot of the maximum wave vector
 for
dynamical stability as a function of $V_0$ and $nU_0$ as solid lines, and
the
corresponding contours for energetic stability are shown as dotted lines.
The numerical calculations in Fig.~\ref{DynamicStab} are converged to
within the width of the contour lines for $\nu_{\rm max}=3$ and $ l_{\rm
max}=3$. For given $V_0$ and $nU_0$, the maximum wave vector for dynamical
stability is always greater than that for energetic stability.
The contours of the maximum wave vector for dynamical stability
are nearly linear for the ranges of
$V_0$ and $nU_0$ investigated. A comparison of the contours for energetic
and dynamical stability in
Fig.~\ref{DynamicStab} shows that the stability boundaries almost coincide
for larger values of $V_0$.

\section{Discussion and conclusions}\label{dis}

Our analytical and numerical calculations show that the band structure of
a
Bose condensate in a one-dimensional optical lattice is
affected dramatically by the presence of interactions between particles.
The appearance of swallow-tail-like loop structures is a rather general
phenomenon.
They occur in the lowest band in the vicinity of the zone
boundary, as reported earlier~\cite{Diakonov2002}, and also, as predicted
in
this paper, near the zone center in higher bands. Indeed, at band gaps at
the
zone boundary or at the zone center one expects such structures to appear
quite
generally on
the band with lower energy if the effective interaction between particles
is
repulsive, the case studied in this paper. For an attractive effective
interaction (negative scattering
length), the loop structures
would appear on the upper band at the gap. While macroscopic condensates
with
negative scattering length are unstable to collapse if the transverse
extent of
the cloud is large, it might be possible to investigate phenomena
associated with
swallow tails in such condensates if the condensate is tightly confined in
the
transverse directions.

 Analytic results were derived using an approximate
wave function containing either two plane waves (for states with wave
vector
close to the zone boundary)
or three plane waves (for states with wave vectors close to zero).
The coupling between plane wave components increases with the
strengths of the interatomic interaction and of
the potential $V_0$, and consequently, the approximate wave functions
become
less accurate. The
potential
induces couplings between plane waves whose wave vectors differ by only
the
smallest
reciprocal lattice vectors $\pm 2\pi/d$, while the interaction term
introduces
couplings
which are less restricted. The simple analytic expressions for the energy
spectra are in good qualitative and, in some cases, quantitative agreement
with the numerical
results. At the zone boundary the approximate
wave function coincides with the exact one, which accounts for the good
agreement with
the numerical results.

For swallow tails to appear, the
interparticle interaction must exceed a critical value which depends on
the
band in question.
We have derived a simple analytic expression, Eq.~(\ref{widthformula}),
for
the width of the
swallow tail around the zone boundary as a function of the interaction
$nU_0$ and the potential $V_0$. According to the
analytical models we have used, in the limit of
a vanishing potential ($V_0 \rightarrow 0$) the width of the swallow tail
at
the
zone center
behaves as $w=nU_0\pi/4E_0d$ and that at the zone boundary as
$w=nU_0\pi/2E_0d$.

The physical interpretation of states on the upper edge of swallow tails
is that they are periodic solitons. These states exist even in the absence
of the lattice potential, and this accounts for the fact that the width of
the swallow tails does not vanish when the lattice potential is absent.

With respect to experimental observability, an important question is
whether or not the stationary states are stable.  For conditions under
which one does not expect swallow-tail structures in the lowest band, the
stability of states has been explored previously by Wu and Niu \cite{WN2},
and for large interactions
but only at the zone boundary in Ref.~\cite{WDN}.
In the present work we
have explored the stability of states on the swallow tail.  States
associated with the upper edge
of the swallow tail are always energetically unstable, since they
correspond to a saddle point in the energy landscape.
The energetic and dynamical stability of the states corresponding to the
lower edge of the swallow tail have been
studied
numerically, and we found that they become more stable as the strength of
the
interatomic interaction increases. We have not calculated growth times for
unstable modes, but the calculations in Ref.\ \cite{WN2} indicate that
these
are
short compared with typical experimental times except under conditions
very
close to the threshold for instability.  Thus we expect growth of
instabilities
to be an important effect in limiting the conditions under
which
states corresponding to the swallow tail may be investigated
experimentally.

The general formalism for studying stability is rather cumbersome, but we
showed in Sec.~\ref{LWanalysis} that for long-wavelength modes one may
develop
a hydrodynamic approach. The detailed numerical calculations indicate that
energetic stability sets in at long wavelengths and therefore the
properties of long-wavelength modes is of direct relevance for determining
the limit of stability of a condensate. Using the hydrodynamic formalism,
we derived a simple analytic expression for the stability limit of the
states at the zone boundary, Eq.~(\ref{AMstablimit}).

We have restricted our present study to stationary states
in which the particle density is periodic with a period equal to the
spacing $d$ of the lattice.
 However, as
will
be discussed elsewhere, stationary states with longer period, e.g.
$2d$ or $3d$, exist.
An example of a state with a particle density which has a period of two
lattice
spacings is a periodic soliton state, with one dark soliton for every two
lattice cells.  The difference in phase between two points separated by
two lattice spacings, $\phi(x+2d)-\phi(x)=\pm \pi$, and therefore the wave
vector is $\pm\pi/2d$.

In order to observe states corresponding to the lower edge of the swallow
tail,
it is desirable that the states be stable.  To achieve this would require
the
mean-field energy $nU_0$ to be about an order of magnitude larger than in
current experiments. The criterion for energetic stability at $k=\pi/d$ is
$nU_0>2E_0$,
corresponding to a a chemical potential $\mu=2E_0$ in the lowest band at
the zone center, while in the experiment of Cataliotti {\it et
al.}~\cite{firenze2} the chemical potential was $\mu\approx 0.2 E_0$.

Throughout our calculations we have assumed that the system is homogeneous
in
the directions transverse to the optical lattice. However, in actual
experiments there is usually a confining potential in these directions.
With
sufficiently tight confinement in the transverse directions, a condensate
is
expected to behave quasi-one-dimensionally, as described in Refs.\
\cite{kavoulakis, modugno}. Such condensates could provide suitable
systems
for
observing some of the non-linear effects predicted in this paper.

\begin{acknowledgments}
It is a pleasure to thank Lars Melwyn Jensen for a number of
useful discussions. The work of M.M. was supported by the Carlsberg
Foundation. We are grateful to Georg Bruun and Lincoln Carr for
helpful discussions
that led to the picture presented in Sec.~\ref{soliton}.
\end{acknowledgments}

\bibliography{mette}

\begin{thebibliography}{24}
\expandafter\ifx\csname natexlab\endcsname\relax\def\natexlab#1{#1}\fi
\expandafter\ifx\csname bibnamefont\endcsname\relax
  \def\bibnamefont#1{#1}\fi
\expandafter\ifx\csname bibfnamefont\endcsname\relax
  \def\bibfnamefont#1{#1}\fi
\expandafter\ifx\csname citenamefont\endcsname\relax
  \def\citenamefont#1{#1}\fi
\expandafter\ifx\csname url\endcsname\relax
  \def\url#1{\texttt{#1}}\fi
\expandafter\ifx\csname urlprefix\endcsname\relax\def\urlprefix{URL }\fi
\providecommand{\bibinfo}[2]{#2}
\providecommand{\eprint}[2][]{\url{#2}}

\bibitem[{\citenamefont{{Ben Dahan} et~al.}(1996)\citenamefont{{Ben Dahan},
  Peik, Reichel, Castin, and Salomon}}]{salomon}
\bibinfo{author}{\bibfnamefont{M.}~\bibnamefont{{Ben Dahan}}},
  \bibinfo{author}{\bibfnamefont{E.}~\bibnamefont{Peik}},
  \bibinfo{author}{\bibfnamefont{J.}~\bibnamefont{Reichel}},
  \bibinfo{author}{\bibfnamefont{Y.}~\bibnamefont{Castin}}, \bibnamefont{and}
  \bibinfo{author}{\bibfnamefont{C.}~\bibnamefont{Salomon}},
  \bibinfo{journal}{Phys. Rev. Lett.} \textbf{\bibinfo{volume}{76}},
  \bibinfo{pages}{4508} (\bibinfo{year}{1996}).

\bibitem[{\citenamefont{Anderson and Kasevich}(1998)}]{kasevich}
\bibinfo{author}{\bibfnamefont{B.~P.} \bibnamefont{Anderson}} \bibnamefont{and}
  \bibinfo{author}{\bibfnamefont{M.~A.} \bibnamefont{Kasevich}},
  \bibinfo{journal}{Science} \textbf{\bibinfo{volume}{282}},
  \bibinfo{pages}{1686} (\bibinfo{year}{1998}).

\bibitem[{\citenamefont{Burger et~al.}(2001)\citenamefont{Burger, Cataliotti,
  Fort, Minardi, Inguscio, Chiofalo, and Tosi}}]{firenze1}
\bibinfo{author}{\bibfnamefont{S.}~\bibnamefont{Burger}},
  \bibinfo{author}{\bibfnamefont{F.~S.} \bibnamefont{Cataliotti}},
  \bibinfo{author}{\bibfnamefont{C.}~\bibnamefont{Fort}},
  \bibinfo{author}{\bibfnamefont{F.}~\bibnamefont{Minardi}},
  \bibinfo{author}{\bibfnamefont{M.}~\bibnamefont{Inguscio}},
  \bibinfo{author}{\bibfnamefont{M.~L.} \bibnamefont{Chiofalo}},
  \bibnamefont{and} \bibinfo{author}{\bibfnamefont{M.~P.} \bibnamefont{Tosi}},
  \bibinfo{journal}{Phys. Rev. Lett.} \textbf{\bibinfo{volume}{86}},
  \bibinfo{pages}{4447} (\bibinfo{year}{2001}).

\bibitem[{\citenamefont{Cataliotti et~al.}(2001)\citenamefont{Cataliotti,
  Burger, Fort, Maddaloni, Minardi, Trombettoni, Smerzi, and
  Inguscio}}]{firenze2}
\bibinfo{author}{\bibfnamefont{F.~S.} \bibnamefont{Cataliotti}},
  \bibinfo{author}{\bibfnamefont{S.}~\bibnamefont{Burger}},
  \bibinfo{author}{\bibfnamefont{C.}~\bibnamefont{Fort}},
  \bibinfo{author}{\bibfnamefont{P.}~\bibnamefont{Maddaloni}},
  \bibinfo{author}{\bibfnamefont{F.}~\bibnamefont{Minardi}},
  \bibinfo{author}{\bibfnamefont{A.}~\bibnamefont{Trombettoni}},
  \bibinfo{author}{\bibfnamefont{A.}~\bibnamefont{Smerzi}}, \bibnamefont{and}
  \bibinfo{author}{\bibfnamefont{M.}~\bibnamefont{Inguscio}},
  \bibinfo{journal}{Science} \textbf{\bibinfo{volume}{293}},
  \bibinfo{pages}{843} (\bibinfo{year}{2001}).

\bibitem[{\citenamefont{Morsch et~al.}(2001)\citenamefont{Morsch, M{\"u}ller,
  Cristiani, Ciampini, and Arimondo}}]{pisa}
\bibinfo{author}{\bibfnamefont{O.}~\bibnamefont{Morsch}},
  \bibinfo{author}{\bibfnamefont{J.~H.} \bibnamefont{M{\"u}ller}},
  \bibinfo{author}{\bibfnamefont{M.}~\bibnamefont{Cristiani}},
  \bibinfo{author}{\bibfnamefont{D.}~\bibnamefont{Ciampini}}, \bibnamefont{and}
  \bibinfo{author}{\bibfnamefont{E.}~\bibnamefont{Arimondo}},
  \bibinfo{journal}{Phys. Rev. Lett.} \textbf{\bibinfo{volume}{87}},
  \bibinfo{pages}{140402} (\bibinfo{year}{2001}).

\bibitem[{\citenamefont{Greiner et~al.}(2002)\citenamefont{Greiner, Mandel,
  Esslinger, H{\"a}nsch, and Bloch}}]{munich}
\bibinfo{author}{\bibfnamefont{M.}~\bibnamefont{Greiner}},
  \bibinfo{author}{\bibfnamefont{O.}~\bibnamefont{Mandel}},
  \bibinfo{author}{\bibfnamefont{T.}~\bibnamefont{Esslinger}},
  \bibinfo{author}{\bibfnamefont{T.~W.} \bibnamefont{H{\"a}nsch}},
  \bibnamefont{and} \bibinfo{author}{\bibfnamefont{I.}~\bibnamefont{Bloch}},
  \bibinfo{journal}{Nature} \textbf{\bibinfo{volume}{415}}, \bibinfo{pages}{39}
  (\bibinfo{year}{2002}).

\bibitem[{\citenamefont{Berg-S{\o}rensen and M{\o}lmer}(1998)}]{kirstine}
\bibinfo{author}{\bibfnamefont{K.}~\bibnamefont{Berg-S{\o}rensen}}
  \bibnamefont{and}
  \bibinfo{author}{\bibfnamefont{K.}~\bibnamefont{M{\o}lmer}},
  \bibinfo{journal}{Phys. Rev. A} \textbf{\bibinfo{volume}{58}},
  \bibinfo{pages}{1480} (\bibinfo{year}{1998}).

\bibitem[{\citenamefont{Choi and Niu}(1999)}]{choi}
\bibinfo{author}{\bibfnamefont{D.-I.} \bibnamefont{Choi}} \bibnamefont{and}
  \bibinfo{author}{\bibfnamefont{Q.}~\bibnamefont{Niu}},
  \bibinfo{journal}{Phys. Rev. Lett.} \textbf{\bibinfo{volume}{82}},
  \bibinfo{pages}{2022} (\bibinfo{year}{1999}).

\bibitem[{\citenamefont{Wu and Niu}(2000)}]{WN1}
\bibinfo{author}{\bibfnamefont{B.}~\bibnamefont{Wu}} \bibnamefont{and}
  \bibinfo{author}{\bibfnamefont{Q.}~\bibnamefont{Niu}},
  \bibinfo{journal}{Phys. Rev. A} \textbf{\bibinfo{volume}{61}},
  \bibinfo{pages}{023402} (\bibinfo{year}{2000}).

\bibitem[{\citenamefont{Wu and Niu}(2001)}]{WN2}
\bibinfo{author}{\bibfnamefont{B.}~\bibnamefont{Wu}} \bibnamefont{and}
  \bibinfo{author}{\bibfnamefont{Q.}~\bibnamefont{Niu}},
  \bibinfo{journal}{Phys. Rev. A} \textbf{\bibinfo{volume}{64}},
  \bibinfo{pages}{061603} (\bibinfo{year}{2001}).

\bibitem[{\citenamefont{Wu et~al.}(2002)\citenamefont{Wu, Diener, and
  Niu}}]{WDN}
\bibinfo{author}{\bibfnamefont{B.}~\bibnamefont{Wu}},
  \bibinfo{author}{\bibfnamefont{R.~B.} \bibnamefont{Diener}},
  \bibnamefont{and} \bibinfo{author}{\bibfnamefont{Q.}~\bibnamefont{Niu}},
  \bibinfo{journal}{Phys. Rev. A} \textbf{\bibinfo{volume}{65}},
  \bibinfo{pages}{025601} (\bibinfo{year}{2002}).

\bibitem[{\citenamefont{Bronski
  et~al.}(2001{\natexlab{a}})\citenamefont{Bronski, Carr, Deconinck, and
  Kutz}}]{bronski1}
\bibinfo{author}{\bibfnamefont{J.~C.} \bibnamefont{Bronski}},
  \bibinfo{author}{\bibfnamefont{L.~D.} \bibnamefont{Carr}},
  \bibinfo{author}{\bibfnamefont{B.}~\bibnamefont{Deconinck}},
  \bibnamefont{and} \bibinfo{author}{\bibfnamefont{J.~N.} \bibnamefont{Kutz}},
  \bibinfo{journal}{Phys. Rev. Lett.} \textbf{\bibinfo{volume}{86}},
  \bibinfo{pages}{1402} (\bibinfo{year}{2001}{\natexlab{a}}).

\bibitem[{\citenamefont{Bronski
  et~al.}(2001{\natexlab{b}})\citenamefont{Bronski, Carr, Deconinck, Kutz, and
  Promislow}}]{bronski2}
\bibinfo{author}{\bibfnamefont{J.~C.} \bibnamefont{Bronski}},
  \bibinfo{author}{\bibfnamefont{L.~D.} \bibnamefont{Carr}},
  \bibinfo{author}{\bibfnamefont{B.}~\bibnamefont{Deconinck}},
  \bibinfo{author}{\bibfnamefont{J.~N.} \bibnamefont{Kutz}}, \bibnamefont{and}
  \bibinfo{author}{\bibfnamefont{K.}~\bibnamefont{Promislow}},
  \bibinfo{journal}{Phys.\ Rev.\ E} \textbf{\bibinfo{volume}{63}},
  \bibinfo{pages}{036612} (\bibinfo{year}{2001}{\natexlab{b}}).

\bibitem[{\citenamefont{Diakonov et~al.}(2002)\citenamefont{Diakonov, Jensen,
  Pethick, and Smith}}]{Diakonov2002}
\bibinfo{author}{\bibfnamefont{D.}~\bibnamefont{Diakonov}},
  \bibinfo{author}{\bibfnamefont{L.~M.} \bibnamefont{Jensen}},
  \bibinfo{author}{\bibfnamefont{C.~J.} \bibnamefont{Pethick}},
  \bibnamefont{and} \bibinfo{author}{\bibfnamefont{H.}~\bibnamefont{Smith}},
  \bibinfo{journal}{Phys. Rev. A} \textbf{\bibinfo{volume}{66}},
  \bibinfo{pages}{013604} (\bibinfo{year}{2002}).

\bibitem[{\citenamefont{Tsuzuki}(1971)}]{tsuzuki}
\bibinfo{author}{\bibfnamefont{T.}~\bibnamefont{Tsuzuki}}, \bibinfo{journal}{J.
  Low Temp. Phys.} \textbf{\bibinfo{volume}{4}}, \bibinfo{pages}{441}
  (\bibinfo{year}{1971}).

\bibitem[{\citenamefont{Kr{\"a}mer et~al.}(2002)\citenamefont{Kr{\"a}mer,
  Pitaevskii, and Stringari}}]{kraemer}
\bibinfo{author}{\bibfnamefont{M.}~\bibnamefont{Kr{\"a}mer}},
  \bibinfo{author}{\bibfnamefont{L.}~\bibnamefont{Pitaevskii}},
  \bibnamefont{and}
  \bibinfo{author}{\bibfnamefont{S.}~\bibnamefont{Stringari}},
  \bibinfo{journal}{Phys. Rev. Lett.} \textbf{\bibinfo{volume}{88}},
  \bibinfo{pages}{180404} (\bibinfo{year}{2002}).

\bibitem[{\citenamefont{Trombettoni and Smerzi}(2001)}]{TS}
\bibinfo{author}{\bibfnamefont{A.}~\bibnamefont{Trombettoni}} \bibnamefont{and}
  \bibinfo{author}{\bibfnamefont{A.}~\bibnamefont{Smerzi}},
  \bibinfo{journal}{Phys. Rev. Lett.} \textbf{\bibinfo{volume}{86}},
  \bibinfo{pages}{2353} (\bibinfo{year}{2001}).

\bibitem[{\citenamefont{Milburn et~al.}(1997)\citenamefont{Milburn, Corney,
  Wright, and Walls}}]{milburn}
\bibinfo{author}{\bibfnamefont{G.~J.} \bibnamefont{Milburn}},
  \bibinfo{author}{\bibfnamefont{J.}~\bibnamefont{Corney}},
  \bibinfo{author}{\bibfnamefont{E.~M.} \bibnamefont{Wright}},
  \bibnamefont{and} \bibinfo{author}{\bibfnamefont{D.~F.} \bibnamefont{Walls}},
  \bibinfo{journal}{Phys. Rev. A} \textbf{\bibinfo{volume}{55}},
  \bibinfo{pages}{4318} (\bibinfo{year}{1997}).

\bibitem[{\citenamefont{Smerzi et~al.}(1997)\citenamefont{Smerzi, Fantoni,
  Giovanazzi, and Shenoy}}]{smerzi}
\bibinfo{author}{\bibfnamefont{A.}~\bibnamefont{Smerzi}},
  \bibinfo{author}{\bibfnamefont{S.}~\bibnamefont{Fantoni}},
  \bibinfo{author}{\bibfnamefont{S.}~\bibnamefont{Giovanazzi}},
  \bibnamefont{and} \bibinfo{author}{\bibfnamefont{S.}~\bibnamefont{Shenoy}},
  \bibinfo{journal}{Phys. Rev. Lett.} \textbf{\bibinfo{volume}{79}},
  \bibinfo{pages}{4950} (\bibinfo{year}{1997}).

\bibitem[{\citenamefont{Coullet and Vandenberghe}(2002)}]{coullet}
\bibinfo{author}{\bibfnamefont{P.}~\bibnamefont{Coullet}} \bibnamefont{and}
  \bibinfo{author}{\bibfnamefont{N.}~\bibnamefont{Vandenberghe}},
  \bibinfo{journal}{J. Phys. B} \textbf{\bibinfo{volume}{35}},
  \bibinfo{pages}{1593} (\bibinfo{year}{2002}).

\bibitem[{\citenamefont{Mahmud et~al.}(2002)\citenamefont{Mahmud, Kutz, and
  Reinhardt}}]{mahmud}
\bibinfo{author}{\bibfnamefont{K.~W.} \bibnamefont{Mahmud}},
  \bibinfo{author}{\bibfnamefont{J.~N.} \bibnamefont{Kutz}}, \bibnamefont{and}
  \bibinfo{author}{\bibfnamefont{W.~P.} \bibnamefont{Reinhardt}},
  \bibinfo{journal}{Phys. Rev. A} \textbf{\bibinfo{volume}{66}},
  \bibinfo{pages}{063607} (\bibinfo{year}{2002}).

\bibitem[{\citenamefont{Lifshitz and Pitaevskii}(1980)}]{landau}
\bibinfo{author}{\bibfnamefont{E.~M.} \bibnamefont{Lifshitz}} \bibnamefont{and}
  \bibinfo{author}{\bibfnamefont{L.~P.} \bibnamefont{Pitaevskii}},
  \emph{\bibinfo{title}{Statistical Physics}} (\bibinfo{publisher}{Pergamon
  Oxford}, \bibinfo{year}{1980}), vol.~\bibinfo{volume}{II}, \bibinfo{note}{\S
  23}.

\bibitem[{\citenamefont{Jackson et~al.}(1998)\citenamefont{Jackson, Kavoulakis,
  and Pethick}}]{kavoulakis}
\bibinfo{author}{\bibfnamefont{A.~D.} \bibnamefont{Jackson}},
  \bibinfo{author}{\bibfnamefont{G.~M.} \bibnamefont{Kavoulakis}},
  \bibnamefont{and} \bibinfo{author}{\bibfnamefont{C.~J.}
  \bibnamefont{Pethick}}, \bibinfo{journal}{Phys. Rev. A}
  \textbf{\bibinfo{volume}{58}}, \bibinfo{pages}{2417} (\bibinfo{year}{1998}).

\bibitem[{\citenamefont{Massignan and Modugno}()}]{modugno}
\bibinfo{author}{\bibfnamefont{P.}~\bibnamefont{Massignan}} \bibnamefont{and}
  \bibinfo{author}{\bibfnamefont{M.}~\bibnamefont{Modugno}},
  \eprint{cond-mat/0205516}.

\end{thebibliography}

\end{document}